\pdfoutput=1
\newcommand*{\ATLASLATEXPATH}{}
\documentclass[cernpreprint,texlive=2011,txfonts,UKenglish]{\ATLASLATEXPATH atlasdoc} 
\usepackage{\ATLASLATEXPATH atlaspackage}
\usepackage{\ATLASLATEXPATH atlasbiblatex}
 
\usepackage{\ATLASLATEXPATH atlascontribute}

\usepackage{\ATLASLATEXPATH atlasphysics}

\graphicspath{{logos/}{figures/}}

\usepackage{EXOT-2014-20_paper-defs}

\newcommand{\jimmy}     {\textsc{JIMMY}\xspace}

\newcommand{\Mcatnlo}   {\textsc{MC@NLO}\xspace}
\newcommand{\Acermc}    {\textsc{AcerMC}\xspace}
\newcommand{\Sherpa}    {\textsc{SHERPA}\xspace}

\newcommand{\Madgraph}  {\textsc{MadGraph}\xspace}
\newcommand{\Pythia}    {\textsc{PYTHIA}\xspace}
\newcommand{\pythiaeight}   {\textsc{Pythia8}\xspace}
\newcommand{\Powheg}    {\textsc{POWHEG}\xspace}
\newcommand{\PowhegBox} {\Powheg-BOX\xspace}

\newcommand{\Perugia}   {\textsc{PERUGIA}\xspace}

\newcommand{\Etmiss}   {\ensuremath{{E}_{\mathrm{T}}^{\mathrm{miss}}}}

\newcommand{\DeltaR}     {\ensuremath{\Delta R}}

\usepackage{multirow}

\AtlasTitle{Search for dark matter produced in association with a Higgs boson decaying to two bottom quarks in $pp$ collisions at $\sqrt{s} = 8$~\TeV~with the ATLAS detector}

\author{ATLAS Collaboration}

\PreprintIdNumber{CERN-PH-EP-2015-247}

\AtlasJournal{Phys.\ Rev.\ D.}

\AtlasAbstract{This article reports on a search for dark matter pair production in association with a Higgs boson decaying to a pair of bottom quarks, using data from 20.3\,fb$^{-1}$ of $pp$ collisions at a center-of-mass energy of $8$~\TeV~collected by the ATLAS detector at the LHC.
The decay of the Higgs boson is reconstructed as a high-momentum $b\bar{b}$~system with either a pair of small-radius jets, or a single large-radius jet with substructure. 
The observed data are found to be consistent with the expected Standard Model backgrounds.
Model-independent upper limits are placed on the visible cross-sections for events with a Higgs boson decaying into $b\bar{b}$~and large missing transverse momentum with thresholds ranging from $150$~\GeV~to $400$~\GeV.
Results are interpreted using a simplified model with a $Z^\prime$ gauge boson decaying into different Higgs bosons predicted in a two-Higgs-doublet model, of which the heavy pseudoscalar Higgs decays into a pair of dark matter particles. Exclusion limits are also presented for the mass scales of various effective field theory operators that describe the interaction between dark matter particles and the Higgs boson.
}

\hypersetup{pdftitle={ATLAS draft},pdfauthor={The ATLAS Collaboration}}

\begin{document}

\maketitle

\section{Introduction}
\label{sec:introduction}
Although dark matter (DM) contributes a large component of the mass-energy of the universe, its properties and interactions with known particles remain unknown~\cite{Bertone:2004pz}. In light of this unsolved puzzle, searches for DM pair-produced at collider experiments provide important information complementary to direct and indirect detection experiments in order to determine whether a signal observed experimentally indeed stems from DM \cite{Bauer:2013ihz}.

The leading hypothesis suggests that most of the DM is in the form of stable, electrically neutral, massive particles, i.e., Weakly Interacting Massive Particles~\cite{Steigman:1984ac}. This scenario gives rise to a potential signature at a proton-proton collider where one or more Standard Model (SM) particles, ``$X$'',  is produced and detected, recoiling against missing transverse momentum (with magnitude \MET) associated with the non-interacting DM. Recent searches at the Large Hadron Collider (LHC) consider ``$X$'' to be a hadronic jet~\cite{Aad:2015zva, Khachatryan:2014rra}, heavy-flavor jet~\cite{Aad:2014vea, Khachatryan:2014uma}, photon~\cite{Aad:2014tda,Chatrchyan:2012tea}, or $W/Z$ boson~\cite{Aad:2014vka,Aad:2013oja}.
The discovery of the Higgs boson $h$~\cite{Aad:2012tfa,Chatrchyan:2012ufa} provides a new opportunity to search for DM production via the $h+\MET$ signature~\cite{Carpenter:2013xra,Berlin:2014cfa,Aad:2015yga}.
In contrast to most of the aforementioned probes, the visible Higgs boson is unlikely to have been radiated from an initial-state quark or gluon, and the signal would give insight into the structure of DM coupling to SM particles. 

Two approaches are commonly used to model generic processes yielding a final state with a particle $X$ recoiling against a system of non-interacting particles. One option is to use non-renormalizable operators in an effective field theory (EFT) framework~\cite{Abercrombie:2015wmb}, where particles that mediate the interactions between DM and SM particles are too heavy to be produced directly in the experiment and are described by contact operators. Alternatively, simplified models that are characterized by a minimal number of renormalizable interactions, and hence explicitly include the particles at higher masses, can be used~\cite{Abdallah:2015ter}. The EFT approach is more model-independent, but is not valid when a typical momentum transfer of the process approaches the energy scale of the contact operators that describe the interaction. Simplified models do not suffer from these concerns, but include more assumptions by design and are therefore less generic. The two approaches are thus complementary and both are included in this analysis.


\section{Signal models and analysis strategy}
\label{sec:method}
\begin{figure}[!ht]
\begin{center}
\subfloat[ EFT ]{\includegraphics[width=0.4\textwidth]{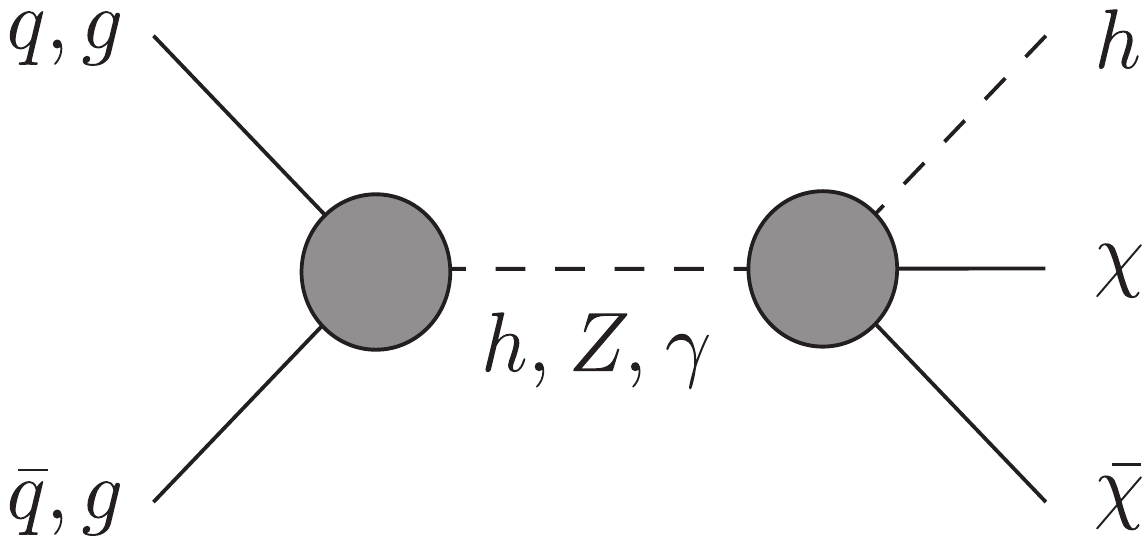}\label{fig:eft}}
\subfloat[ \Zprime-2HDM ]{\includegraphics[width=0.45\textwidth]{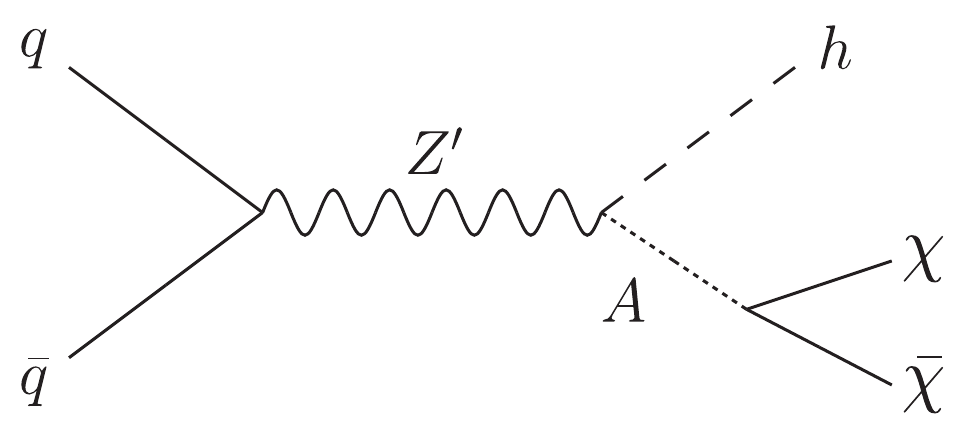}\label{fig:zp2hdm}} 
\end{center}
\caption{Feynman diagrams for (a) the EFT and (b) the \Zprime-2HDM models. The $\chi$ is the DM particle. The $h$ is the $125$~\GeV~observed Higgs boson. In (a), the left dark circle denotes the coupling from $q\bar{q}$ or $gg$ to an electroweak boson ($h$, $Z$, $\gamma$) that mediates the DM+$h$ production, and the right dark circle represents the contact operator in the EFT framework between DM, the Higgs boson, and the mediator. In (b), the $A$ is the heavy pseudoscalar in the two-Higgs-doublet model. }
\label{fig:Feynman}
\end{figure}

Using the EFT approach, a set of models described by effective operators at different dimensions is considered, as shown in Figure~\ref{fig:Feynman}\subref{fig:eft}.
Following the notation in Ref.~\cite{Carpenter:2013xra}, the effective operators in ascending order of their dimensions are:

\begin{align}
& \lambda|\chi|^2|H|^2 & \textrm{(Scalar DM, dimension-4)} \label{eq:dim4}\\
& \frac{1}{\Lambda}\bar{\chi}i\gamma_5\chi|H|^2 & \textrm{(Fermionic DM, dimension-5)} \label{eq:dim5}\\
& \frac{1}{\Lambda^2}\chi^\dagger \partial^\mu \chi H^\dagger D_\mu H & \textrm{(Scalar DM, dimension-6)} \label{eq:dim6}\\
& \frac{1}{\Lambda^4}\bar \chi \gamma^\mu \chi  B_{\mu\nu} H^\dagger D^\nu H & \textrm{(Fermionic DM, dimension-8)} \label{eq:dim8} 
\end{align}

Here $\chi$ is the DM particle, which is a gauge singlet under $SU(3)_C \times SU(2)_L \times U(1)_Y$ and may be a scalar or a fermion as specified, $D_\mu(^\nu)$ is the covariant derivative for the full gauge group, and $B_{\mu\nu}$ is the $U(1)_Y$ field strength tensor.
The parameters of these models are the DM particle mass $m_\chi$, and the coupling parameter $\lambda$ or the suppression scale $\Lambda$
of the heavy mediator that is not directly produced but described by a contact operator in the EFT framework.

A simplified model is also considered which contains a \Zprime\ gauge boson and two Higgs fields resulting in five Higgs bosons (often called the two-Higgs-doublet model, 2HDM)~\cite{Berlin:2014cfa}, where the DM particle is coupled to the heavy pseudoscalar Higgs boson $A$, as shown in Figure~\ref{fig:Feynman}\subref{fig:zp2hdm}. In this model (\Zprime-2HDM), the \Zprime\ boson is produced resonantly and decays into $h$ and $A$ in a Type $2$ two-Higgs-doublet model~\cite{Branco:2011iw}, where $h$ is the scalar corresponding to the observed Higgs boson, and $A$ has a large branching ratio to DM. The \Zprime\ boson can also decay to a Higgs boson and a $Z$ boson, which in turn decays to a pair of neutrinos, thus mimicking the expected signature. While the $Ah$ decay mode is dominant for most of the parameter space probed in this analysis, the $Zh$ decay mode is an important source of signal events at large $\tan{\beta}$ (the ratio of the vacuum expectation values for the two-Higgs-doublets).  Both sources of a Higgs boson plus missing transverse momentum are included for the analysis of this model. The results presented are for the alignment limit, in which the scalar Higgs mixing angle $\alpha$ is related to $\beta$ by $\alpha = \beta - \pi/2$. Only regions of parameter space consistent with precision electroweak constraints on the $\rho_0$ parameter~\cite{Beringer:1900zz} and with constraints from direct searches for dijet resonances~\cite{Aaltonen:2008dn, CMS:2012yf, Khachatryan:2015sja} are considered. The \Zprime~boson does not couple to leptons in this model, avoiding potentially stringent constraints from dilepton searches. As the $A$ boson is produced on-shell and decays into DM, the mass of the DM particle does not affect the kinematic properties or cross-section of the signal process when it is below half of the $A$ boson mass. Hence, the \Zprime-2HDM model is interpreted in the parameter spaces of \Zprime\ mass ($m_{\Zprime}$), $A$ mass ($m_A$), and $\tan{\beta}$, with the \Zprime\ gauge coupling fixed to its $95\%$ confidence level (CL) upper limit per \Zprime\ mass and $\tan{\beta}$ value from the aforementioned electroweak and dijet search constraints. 

This article describes the search for DM pair production in association with a Higgs boson using the full 2012 ATLAS data set corresponding to 20.3 fb$^{-1}$ of $pp$ collisions with center-of-mass energy $\sqrt{s}=8$ TeV. 
The final state is a Higgs boson decaying to a pair of bottom quarks and large missing transverse momentum. 
Two Higgs boson reconstruction techniques are presented that are complementary in their acceptance. The first, ``resolved'' technique reconstructs Higgs boson candidates from pairs of nearby \akt jets~\cite{Cacciari:2008gp} each reconstructed with radius parameter $R = 0.4$ and each identified as having a $b$-hadron within the jet using a multivariate $b$-tagging algorithm \cite{Aad:2015ydr}. This resolved technique offers good efficiency over a wide kinematic range with the Higgs boson transverse momentum \pt between 150 and 450\ GeV. 
However, for a Higgs boson with $\pt \gtrsim 450$~\GeV, the high momentum (``boost'') of the Higgs boson causes the two jet cones containing the $b$- and $\bar{b}$-quarks from the Higgs boson decay to significantly overlap, leading to a decrease in the reconstruction efficiency of the two $b$-tagged \akt jets with $R = 0.4$.
This motivates the use of the same  ``boosted'' Higgs boson reconstruction technique in Ref.~\cite{Aad:2015uka}. The acceptance for these higher-\pt Higgs bosons is maintained through the use of the internal structure of jets, known as ``jet substructure'' techniques, and the subjet $b$-tagging algorithms. The Higgs boson candidate is reconstructed as a single \akt $R=1.0$ jet, trimmed~\cite{Krohn2010} with subjet radius parameter $R_{\mathrm{sub}}=0.3$ and subjet transverse momentum fraction ${\pt}_i/\pt^{\mathrm{jet}}<0.05$, where ${\pt}_i$ is the transverse momentum of the $i$-th subjet and $\pt^{\mathrm{jet}}$ is the $\pt$ of the untrimmed jet~\cite{Aad:2013gja, Aad:2015rpa}. This $R=1.0$ jet must be associated with two $b$-tagged \akt $R = 0.3$ jets reconstructed only from charged particle tracks (track-jets)~\cite{ATL-PHYS-PUB-2014-013}.
The use of track-jets with a smaller $R$ parameter allows the decay products of Higgs bosons with higher \pt to be reconstructed.

The interplay between the two sets of models and analysis methods has been studied. In the \Zprime-2HDM simplified model, the resonant production and decay of the \Zprime\ boson leads to clear peaks in the \MET spectra, the positions of which depend on the \Zprime\ and $A$ mass values. In most of the parameter space probed with \Zprime\ mass between 600 and 1400 GeV, and $A$ mass between 300 and 800 GeV (where kinematically allowed), a higher signal sensitivity is achieved in the resolved channel. On the other hand, the EFT models display very different kinematics with wide tails in high \MET extending beyond $450$~\GeV, warranting a ``boosted'' reconstruction of the Higgs boson. Given the clear advantage of one analysis channel over the other for either set of models, and for simplicity, the results for the \Zprime-2HDM model are given using the resolved analysis, and the EFT models are interpreted using the boosted analysis. 

The final signal regions are defined with four increasing thresholds for the missing transverse momentum in the resolved channel, and two thresholds in the boosted channel. To search for the possible presence of non-SM signals, the total numbers of observed events after applying all selection criteria are compared with the total number of expected SM events taking into account their respective uncertainties in both channels. 
Unlike previous ATLAS searches for resonant production with a similar final state~\cite{Aad:2015yza, Aad:2015wra}, this analysis explores different theoretical models, focuses on the fully hadronic channel with data-driven methods to estimate the main backgrounds, and most importantly, applies selections extending to large \MET~ utilizing ``resolved'' as well as ``boosted'' techniques. The approach for extracting limits in this analysis is also more suited for the models considered here, and reduces the theoretical uncertainty from modeling and fitting of the signal shape.

\section{ATLAS detector}
\label{sec:detector}
ATLAS is a multi-purpose particle physics experiment~\cite{Aad:2008zzm} at the LHC. The detector\footnote{ATLAS uses a right-handed coordinate system with its origin at the nominal interaction point (IP) in the center of the
detector and the $z$-axis along the beam pipe. The $x$-axis points from the IP to the center of the LHC ring, and the $y$-axis points
upwards. Cylindrical coordinates $(r, \phi)$ are used in the transverse plane, $\phi$ is the azimuthal
angle around the beam pipe. The pseudorapidity $\eta$ is defined in terms of the polar angle $\theta$ as $\eta = -\ln[\tan (\theta/2)]$.}
consists of inner tracking devices surrounded by a superconducting solenoid, electromagnetic
and hadronic calorimeters, and a muon spectrometer.
The inner tracking system provides charged-particle tracking and vertex reconstruction in the pseudorapidity region of $|\eta| < 2.5$. It consists of a silicon pixel detector, a silicon microstrip
tracker, and a transition radiation tracker. The system is surrounded by a solenoid that produces a 2\,T axial magnetic
field.
The central calorimeter system consists of a liquid-argon electromagnetic sampling calorimeter
with high granularity and a steel/scintillator-tile calorimeter providing hadronic energy measurements in
the central pseudorapidity range ($|\eta| < 1.7$). The endcap and forward regions are instrumented with liquid-argon
calorimeters for electromagnetic and hadronic energy measurements up to $|\eta| = 4.9$. The muon spectrometer
is operated in a magnetic field provided by air-core superconducting toroids and includes tracking
chambers for precise muon momentum measurements up to $|\eta| = 2.7$ and trigger chambers covering the range of $|\eta| < 2.4$.
A three-level trigger system is used to select interesting events~\cite{Aad:2012xs}.
The Level-1 (L1) trigger reduces the event rate to below 75\,kHz using hardware-based trigger
algorithms acting on a subset of detector information. Two levels of software-based triggers, referred to collectively as the High-Level Trigger (HLT),
further reduce the event rate to approximately 400\,Hz using information from the entire detector.

\section{Data and simulation samples}
\label{sec:data-mc}

The data sample used in this analysis, after data quality requirements are applied, corresponds to an integrated luminosity of 20.3 fb$^{-1}$. 
The primary data sample is selected using an \Etmiss\, trigger.   
The L1 \Etmiss\, trigger threshold is $60$~\GeV, and the HLT \Etmiss\, trigger threshold is $80$~\GeV.
The trigger efficiency is above $98\%$ for events passing the full offline selection across the full \Etmiss\, range considered in this analysis.
Muon triggers with transverse momentum thresholds at the HLT of 24 GeV for muons with surrounding inner detector tracking activity below a predefined level, i.e., isolated muons~\cite{combinedmuons}, and 36 GeV for muons with no isolation requirement, are used to select the muon data used for the estimation and validation of backgrounds in the control regions. A photon trigger with a transverse momentum threshold of $120$~\GeV~ at the HLT is used to select events with a high $\pt$ prompt photon for data-driven $Z(\rightarrow\nu\bar{\nu})$+jets background estimation (Section \ref{sec:znnbackground}). 

Monte Carlo (MC) simulated event samples are used to model both the signal and backgrounds. 
Effects of multiple proton--proton interactions (pileup) as a function of the instantaneous luminosity are taken into account by overlaying simulated minimum-bias events generated with \pythiaeight~\cite{pythia8} onto the hard-scattering process, such that the distribution of the average number of interactions per bunch crossing in the MC simulated samples matches that in the data. 
The simulated samples are processed either with a full ATLAS detector simulation~\cite{Aad:2010ah} based on the {\sc Geant4} program~\cite{Geant4}, or a fast simulation of the response of the electromagnetic and hadronic calorimeters~\cite{atlas_simulation}. The results based on fast simulations are validated against fully simulated samples and the difference is found to be negligible. The simulated samples are further processed with a simulation of the trigger system. Both the simulated events and the data are reconstructed and analyzed with the same analysis chain, using the same event selection criteria.

Table \ref{tab:Generators} summarizes the various event generators and parton distribution function (PDF) sets, as well as
parton shower and hadronization software used for the analyses presented in this article.

\begin{table}[htb]
\caption{
  Summary of MC event generators, PDF sets, and parton shower and hadronization models utilized in the analyses for both the signal and background processes. 
}\label{tab:Generators} 

\begin{center}
\small
\begin{tabular}{l l l l}
\toprule
 Model / Process & Generator & PDF & Parton Shower / Hadronization \\ 
\midrule
 \Zprime-2HDM & \Madgraph v1.5.1 & MSTW2008LO  & \Pythia~v8.175 with AU2 tune\\
 EFT models                & \Madgraph v1.5.1 & CTEQ6L1 & \Pythia~v8.175 with AU2 tune\\
 $W/Z/\gamma$+jets & \Sherpa v1.4.3& CT10 &  \Sherpa v1.4.3 \\
 \ttbar & \PowhegBox~v1.0 r2129 & CT10 & \Pythia~v6.427 with P2011C tune \\
 Single top ($s$-ch., $Wt$) & \Mcatnlo v3.31 & CT10 & \jimmy~v4.31 with AUET2 tune \\
 Single top ($t$-ch.) & \Acermc v3.8 & CTEQ6L1 & \Pythia~v6.426 with AUET2B tune \\
$WW/WZ/ZZ$ (resolved) & \Herwig~v6.520  & CTEQ6L1 &  \jimmy~v4.31 with AUET2 tune  \\
$WW/WZ/ZZ$ (boosted) & \Powheg~r2330.3  & CTEQ6L1 &  \Pythia~v8.175 with AU2 tune\\
$q\bar{q}\to Vh$ & \Pythia~v8.175  & CTEQ6L1 & \Pythia~v8.175 with AU2 tune\\
$gg\to Zh$ & \Powheg~r2330.3 & CT10 &  \Pythia~v8.175 with AU2 tune\\
multijet &  \Pythia~v8.160 & CT10 &  \Pythia~v8.160 with AU2 tune\\
\bottomrule
\end{tabular}
\end{center}
\end{table}

Signal samples are generated with \Madgraph~\cite{Alwall:2011uj}, interfaced to \pythiaeight using the AU2 parameter settings (tune)~\cite{AU2} for parton showering, hadronization, and underlying event simulation. The Higgs boson mass is fixed to 125\,GeV. The MSTW2008LO leading-order (LO) PDF set~\cite{Martin:2009nx} is used for the \Zprime-2HDM model, while the CTEQ6L1 PDF set \cite{cteq6l1} is used for the EFT models.  For the \Zprime-2HDM model, samples are produced with \Zprime\ mass values between 600 and 1400 GeV, $A$ mass values between 300 and 800 GeV (where kinematically allowed), and DM mass values between 10 and 200 GeV but always less than half the $A$ mass.  In addition, $\Zprime\rightarrow Zh$ samples are produced for \Zprime\ mass values between 600 and 1400 GeV.
For the EFT models, samples are produced for scalar and fermionic DM particle masses ranging from 1 to 1000~\GeV\ for both $hh$ and $hZ$ coupling to DM.

A variety of samples are used in the background determination.  The dominant $Z(\rightarrow\nu\bar{\nu})$+jets background is determined from data (Section~\ref{sec:znnbackground}), and samples simulated with \Sherpa \cite{Gleisberg:2008ta}  for $Z(\rightarrow\nu\bar{\nu})$+jets, $Z(\to\ell\ell)$+jets, and $\gamma$+jets are also used in the calculation process.   
The $W(\rightarrow \ell\nu)$+jets processes are generated with \Sherpa and are normalized using data as described in Section~\ref{sec:1LepCR}.
All the \Sherpa samples are generated using the CT10 PDF set \cite{CT10}.
The \ttbar background is generated with \PowhegBox~\cite{Powheg} interfaced with \pythiasix and the \Perugia 2011C tune \cite{Skands:2010ak}.  
Single top quark production in the $s$- and $Wt$-channels are produced with \Mcatnlo~\cite{Mcatnlo, Frixione:2002ik, Frixione:2005vw} interfaced with \jimmy~\cite{Jimmy}, while the $t$-channel process is produced with \Acermc \cite{Acermc} interfaced with \pythiasix. 
The Diagram Removal scheme~\cite{Frixione:2008yi} is used in the single top quark production in the $Wt$-channel to remove potential overlap with \ttbar production due to interference of the two processes.
A top quark mass of $172.5$~\GeV~is used consistently. 
The cross-sections of the \ttbar and single-top-quark processes are determined at next-to-next-to-leading order (NNLO) in QCD including resummation of next-to-next-to-leading logarithmic (NNLL) soft gluon terms with Top++2.0~\cite{Cacciari:2011hy, Beneke:2011mq, Baernreuther:2012ws, Czakon:2012pz, Czakon:2012zr, Czakon:2013goa, Czakon:20142930}. 
The normalization and uncertainties are calculated using the PDF4LHC prescription~\cite{Botje:2011sn} with the MSTW2008 $68\%$ CL NNLO~\cite{Martin:2009nx, Martin:2009bu}, CT10 NNLO~\cite{CT10, Gao:2013xoa}, and NNPDF2.3~\cite{Ball:2012cx} PDF sets. 
Additional kinematic-dependent corrections to the \ttbar sample and normalizations  determined from data are described in Section~\ref{sec:1LepCR}.
Diboson ($ZZ$, $WW$, and $WZ$) production is simulated with two different generators, both \Herwig \cite{Herwig} interfaced to \jimmy and \Powheg interfaced to \pythiaeight. The differences in event yield and kinematic distributions between the two simulated samples are found to be minimal in the analyses. 
The diboson samples are normalized to calculations at next-to-leading order (NLO) in QCD performed using MCFM~\cite{MCFM}.
The multijet background is estimated from data (Section~\ref{sec:qcdbackground}), with samples simulated with \pythiaeight used for validation in the control regions. 
For SM production of $Zh$ and $Wh$, $\pythiaeight$ is used with CTEQ6L1 PDFs, and the samples were normalized to total cross-sections calculated at NLO~\cite{Han:1991ia}, and NNLO~\cite{Brein:2003wg} in QCD, respectively, with NLO electroweak corrections~\cite{Ciccolini:2003jy} in both cases.

\section{Object reconstruction}
\label{sec:object}

This analysis requires the reconstruction of  muons, electrons, photons, jets, and missing transverse momentum. Object reconstruction efficiencies in simulated events are corrected to reproduce the performance measured in data, and their systematic uncertainties are detailed in Section~\ref{sec:ResolvedSyst}.

Muon candidates are identified from tracks that are well reconstructed inside both the inner detector and the muon spectrometer~\cite{combinedmuons}. They must fulfill \pT $> 6$~\GeV~and \abseta $< 2.5$ requirements. Furthermore, they are required to satisfy the ``{\it tight}'' muon identification quality criteria~\cite{combinedmuons}.
To reject cosmic-ray muons, muon candidates are required to be consistent with production at
the primary vertex, defined as the vertex\footnote{Proton--proton collision vertices are reconstructed requiring that at least five tracks with \pt~$>$ 0.4\,\GeV~are associated with a given vertex.} with the highest $\Sigma (p_\mathrm{T}^{\mathrm{track}} )^2$, where $p_\mathrm{T}^{\mathrm{track}}$ refers to the transverse momentum of each track. 
In the muon control region or during the overlap removal procedure of the boosted channel, muon candidates are required to be isolated to reduce the multijet background. The scalar sum of the transverse momenta of tracks with $\pT > 1$~\GeV~within a cone of $\Delta R =\sqrt{(\Delta\eta)^2 + (\Delta\phi)^2} = 0.3$ around the muon track excluding the muon (tracking isolation), as well as the transverse energy measured in the calorimeter in a cone of $\Delta R = 0.3$ (excluding the energy lost by the muon itself) around the muon track (calorimeter isolation), is required to be less than 12\% of the muon $\pT$.

Electron candidates are identified as
tracks that are matched to a cluster meeting shower-shape criteria in the electromagnetic calorimeter. Each electron candidate should have \pT $> 7$~\GeV and is within \abseta $< 2.47$.
To suppress contamination from multijet background, electron candidates must satisfy the ``{\it medium++}'' electron shower shape and track selection criteria, based on Ref.~\cite{mediumelectrons} and modified to accommodate the increased pileup in $8$~TeV data. Isolated electrons are used in the boosted channel during the overlap removal procedure. These isolated electrons must meet tracking and calorimeter isolation requirements. The scalar sum of the transverse momenta of tracks with $\pT > 1$~\GeV~within a cone of $\Delta R = 0.3$ around the electron track excluding the electron is required to be less than 16\% of the electron $\pT$. The transverse energy measured in the calorimeter in a cone of $\Delta R = 0.3$ (excluding the energy lost by the electron itself) around the electron track is required to be less than 18\% of the electron $\pT$.

Photon candidates must satisfy the {\it tight} quality criteria with $\pt > 10$~\GeV~and $|\eta| < 2.37$ \cite{Aad:2014nim}. Additionally, the isolated photons used in the $Z(\nu\bar{\nu})$+jets background estimation must have $\pt > 125$~\GeV, and the sum of the energy deposit in the topological calorimeter clusters within a radius $R=0.4$ with respect to the photon direction, but excluding the photon, must be less than 5~\GeV. 

Jets are reconstructed~\cite{Cacciari:2011ma} using the \akt\ jet clustering algorithm from topological clusters of calorimeter cells that are locally calibrated to the hadronic energy scale~\cite{Aad:2014bia}. Small-radius (small-$R$; radius parameter $R = 0.4$) jets as well as large-radius (large-$R$; $R=1.0$) jets are used. 
The effects of pileup on small-$R$ jet energies are accounted for by a correction based on jet area~\cite{Cacciari:2008gn}.
The jet trimming algorithm~\cite{Krohn2010} is applied to the reconstruction of large-$R$ jets to minimize the impact of energy depositions due to pileup and the underlying event. This algorithm reconstructs subjets within the large-$R$ jet using the $k_{\mathrm t}$ algorithm~\cite{Catani:1993hr} with radius parameter $R_{\rm sub} = 0.3$, then removes any subjet with $\pt$ less than 5\% of the large-$R$ jet \pt.
The energies of all jets and the masses of the large-$R$ jets are then calibrated to their values at particle level using \pt- and $\eta$-dependent factors determined from simulation; small-$R$ jets are further calibrated using {\it in situ} measurements~\cite{Aad:2011he}.
Small-$R$ jets with $\pt < 50$~\GeV~and $|\eta|<2.4$ are required to have at least 50\% of the \pt~sum of tracks
matched to the jet belonging to tracks originating from the primary vertex (jet vertex fraction) to suppress the effects of pileup interactions~\cite{jvf}.
Small-$R$ jets are required to satisfy either $\pT > 25$~\GeV~and $|\eta| < 2.4$ or $\pT > 30$~\GeV~and $2.4<|\eta| < 4.5$, while large-$R$ jets are required to satisfy $\pT > 300$~\GeV~and $|\eta| < 2.0$.

Track-jets are built from tracks using the \akt\ algorithm with $R = 0.3$. Tracks are required to satisfy $\pt > 0.5$ \GeV~and $|\eta| < 2.5$,  the transverse and longitudinal impact parameters with respect to the primary vertex below $1.5$~mm, and a set of hit criteria to ensure that those tracks are consistent with
originating from the primary vertex, thereby reducing the effects of pileup.
Track-jets are matched to large-$R$ jets using a process called
``ghost association'' \cite{Cacciari:2007fd,Cacciari:2008gn}.
Track-jets with $\pT > 20$~\GeV~and $|\eta| < 2.5$ are kept for further analysis.

Small-$R$ jets and track-jets containing $b$-hadrons are identified (``$b$-tagged'')
using the properties of the tracks associated with them, the most important being the impact parameter of each track
(defined as the track's distance of closest approach to the primary vertex in the transverse plane), as well as the presence and properties of displaced vertices.
The ``MV1'' $b$-tagging algorithm~\cite{Aad:2015ydr} used in this analysis combines the above information
using a neural network and is configured to achieve an average efficiency of $60\%$ for tagging
small-$R$ jets with $b$-quarks \footnote{In simulation, a jet is labeled as a $b$-quark jet if a $b$-quark (after final-state radiation) with transverse momentum above 5\,\GeV~is identified within a cone of $\Delta R =0.3$ around the jet axis. If no $b$-quark is identified, the jet is labeled as a charm-quark jet if a charm-quark is identified with the same criteria. If no charm quark is identified, the jet is labeled as a $\tau$-jet if a $\tau$-lepton is identified with the same criteria. Otherwise the jet is labeled as a light-flavor jet.}, and has misidentification probabilities of $\sim15\%$ for charm-quark jets and less than $1\%$ for light-flavor jets, as determined in an MC sample of \ttbar~events.
For track-jets, the corresponding numbers are $74\%$ for $b$-quark jets, $15\%$ for charm-quark jets, and $<1.5\%$ for light-flavor jets. 
The $b$-tagging algorithm is trained on MC simulations and its efficiency is scaled to match data based on studies of candidate \ttbar and multijet events~\cite{Aad:2015ydr,Aad:2015uka}.
For charm- and light-flavor track-jets, the efficiency calibrations for the small-$R$ jets are used, with additional uncertainties to account for possible differences in $b$-tagging performance between small-$R$ jets and track-jets.
The flavor-tagging efficiency is only calibrated up to \pt of $300$~\GeV~for $b$- and $c$-tagged small-$R$ jets, $750$~\GeV~for light-flavor-tagged small-$R$ jets, and $250$~\GeV~for $b$-tagged track-jets.
Beyond the maximum \pt, additional uncertainties on the $b$-tagging efficiency are extracted from the last calibrated \pt bin with additional uncertainties based on studies of MC-simulated events with high-\pt jets.

Since each type of object reconstruction proceeds independently, the same calorimeter cells or tracks might be used for multiple physics objects. This can lead to double counting of energy and the dual-usage must be resolved. In addition, two separate but close-by objects can also potentially introduce bias in the reconstruction process. To address the problem of duplication while preserving heavy-flavour jets with semi-leptonic decays or the problem where close-by objects bias each other’s position or energy reconstruction, the following sequential overlap removal procedures are implemented separately for the resolved and the boosted channel. 
In the resolved channel, an object is considered to be an electron (photon) and a small-$R$ jet is discarded if the electron (photon) candidate and the small-$R$ jet that is not $b$-tagged overlap within $\Delta R< 0.2$. If an electron (photon) candidate and any small-$R$ jet have angular separation in the range of $0.2 < \Delta R < 0.4$, or if an electron (photon) candidate and a $b$-tagged small-$R$ jet overlap within $\Delta R< 0.2$ of each other, then the electron (photon) is discarded and the object is considered a small-$R$ jet. 
If a muon candidate and a small-$R$ jet overlap within $\Delta R<0.4$, then the muon is discarded and the small-$R$ jet is retained. 
In the boosted channel, an object is considered to be an electron candidate and a small-$R$ jet is removed if the electorn that is isolated and the small-$R$ jet overlap within $\Delta R<0.2$.  Electron or muon candidates will be removed if they and any small-$R$ jet overlap within $\Delta R<0.4$. Furthermore, large-$R$ jets are eliminated if an isolated photon is found within $\Delta R<1.0$ of the large-$R$ jet. Track-jets are discarded if an isolated electron or an isolated muon is found within $\Delta R<0.1$ of the track-jet.

The missing transverse momentum $\vecMET$ is defined as the negative vector sum of the transverse
momenta of jets, electrons, photons, and topological calorimeter clusters not assigned to any reconstructed objects~\cite{etmissref}. 
The transverse momenta of reconstructed muons are included, with the energy
deposited by these muons in the calorimeters properly removed to avoid double-counting.
In addition, a track-based missing transverse momentum vector $\vecMPT$ is calculated as the 
negative vector sum of the transverse momenta of tracks with $|\eta| < 2.4$ and the transverse and longitudinal impact parameters with respect to the primary vertex below $1.5$~mm. 

\section{Event selection}
\label{sec:eventselection}

A set of common preselection criteria based on objects described in Section~\ref{sec:object} is used for events to be considered for the resolved and boosted channels.
An initial $\Etmiss+\rm{jets}$ sample is obtained by requiring an event to have passed the 80 GeV HLT \MET trigger, to have an offline \Etmiss~$> 100$~\GeV~for the resolved channel (\Etmiss~$> 200$~\GeV~for the boosted channel) and to have at least one small-$R$ jet. No electron, muon and photon candidates should be present in the event. Events must have at least one identified $pp$ collision vertex and be produced in stable beam conditions with all relevant subdetectors functioning properly.  
To suppress contamination from multijet events, the smallest azimuthal angle between $\vecMET$ and small-$R$ jets is required to be greater than 1.0.

\begin{table*}[h]
\begin{center}
\caption{The event selection criteria for signal regions in the resolved and boosted channels. The symbol $j$ represents an \akt jet ($R=0.4$), $j^{\rm trk}$ a track-jet ($R=0.3$), $J$ a trimmed \akt jet ($R=1.0$), $b$ a $b$-tagged \akt jet ($R=0.4$), and $b^{\rm trk}$ a $b$-tagged \akt track-jet ($R=0.3$).  Each $b$-tagged track-jet is matched by ghost association to the leading-$\pt$ large-$R$ jet. The subscript index $i$ of each jet collection means the $i$-th jet in descending order of the transverse momentum, of which $j_i$ are inclusive and may or may not be $b$-tagged. The variable $\Delta\phi_{\textrm{min}}(\vecMET, j_i)$ refers to the smallest $\phi$ angular separation between the $\vecMET$ and any \akt jet ($R=0.4$) in the event. }\label{tab:eventselection}
\begin{tabular}{l|c|c}
\hline \hline
 & Resolved & Boosted \\ \hline
$\Delta\phi_{\textrm{min}}(\vecMET, j_i)$ & $>1.0$  & $> 1.0$ \\ \hline
\multirow{2}{*}{Jet multiplicity} & $2\le n_j \le 3$  & $n_J \ge 1$ \\ & & $n_{j^{\rm trk}}\ge 2$ \\ \hline
$b$-jet (60\% eff.) \pT & $\pt^{b_{1}}>100$~\GeV~ & - \\ \hline
$b$-jet multiplicity  & $n_b \ge 2$  (60\% eff.) & $ n_{b^{\rm trk}}= 2$ (70\% eff.) \\ \hline
\multirow{2}{*}{Jet \pT} & $\pt^{b_2}>60$~\GeV~when $n_j = 3$ &  \multirow{2}{*}{ $\pt^{J_1}>350$~\GeV~} \\ 
                                     & $\pt^{j_2}>100$~\GeV~when $n_j = 3$ & \\ \hline
$\Delta\phi(\vecMET,\vecMPT)$ & - & $< \pi/2$ \\ \hline
Dijet separation &$\Delta R(j_1,j_2) < 1.5$ & - \\ \hline
Invariant mass & 90 GeV $\le m_{b_1 b_2} \le 150$~\GeV~& 90 GeV $\le m_{J_1} \le 150$~\GeV~\\ \hline
\MET & $>$ 150, 200, 300, or 400 GeV & $>$ 300 or 400 GeV \\ \hline
\hline
\end{tabular}
\end{center}
\end{table*}

For the resolved channel, a further set of selection criteria is chosen by optimizing the sensitivity to a simulated \Zprime-2HDM signal in the presence of the expected background.  The selection criteria are summarized in Table~\ref{tab:eventselection}. If no explicit jet \pT threshold is specified that means only the initial selection criteria described previously are required. The requirements on the \pT of the subleading $b$-tagged jet, $\pt^{b_2}$, and that of the subleading jet, $\pt^{j_2}$, for events containing three jets were found to be effective in removing top quark background. The minimum \MET value required increases with $m_{\Zprime}$ to take advantage of the harder \MET spectrum for higher $\Zprime$ mass values. 
The best signal sensitivity at $\tan{\beta}=1$ for the signal samples used in this analysis is achieved by requiring a minimum \MET of $200$~\GeV~for $m_{\Zprime} = 600$~\GeV, $300$~\GeV~for $m_{\Zprime} = 800$~\GeV, and $400$~\GeV~for $m_{\Zprime} = 1000$--$1400$~\GeV.
The product of the detector acceptance and reconstruction efficiency (selection efficiency) of the $\Zprime\to h(b\bar{b})+\MET$ signal after the full set of selection requirements varies from 5\% to 10\% depending on $m_{\Zprime}$ and $m_{A}$. The number of expected signal events after full selection in the \Zprime-2HDM model for a few selected values of $m_{\Zprime}$, $m_{A}$ and $\tan{\beta}$ are shown in Table~\ref{tab:zpsignalyield} for the $\Zprime \to A(\chi\bar{\chi})h(b\bar{b})$ and $\Zprime \to Z(\nu\bar{\nu})h(b\bar{b})$ processes respectively. 

\begin{table*}[h]
  \begin{center}
   \caption{ The number of expected \Zprime-2HDM signal events after full selection for selected points in parameter space. Left to right: values of $m_{\Zprime}$, $m_{A}$, and $\tan{\beta}$, the \MET requirement for the given parameter values, the signal yield from the $\Zprime\to A(\chi\bar{\chi})h(b\bar{b})$ and $\Zprime\to Z(\nu\bar{\nu})h(b\bar{b})$ processes respectively. }
    \label{tab:zpsignalyield}
    \begin{tabular}{llcccccc}
\hline \hline
 & $m_{\Zprime}$  & $m_{A}$ & $\tan{\beta}$ & \MET & $\Zprime\to A(\chi\bar{\chi})h(b\bar{b})$ & $\Zprime\to Z(\nu\bar{\nu})h(b\bar{b})$  \\ \hline
 & $600$~\GeV & $300$~\GeV & $0.3$ &$> 150$~\GeV & $10$ & $1.1$ \\ 
 & $600$~\GeV & $300$~\GeV & $1$ &$> 200$~\GeV & $3.5$ & $11.9$ \\
 & $800$~\GeV & $300$~\GeV & $1$ &$> 300$~\GeV & $10.4$ & $6.8$ \\
 & $1000$~\GeV & $300$~\GeV & $0.3$ &$> 400$~\GeV & $12.2$ & $0.4$ \\ 
 & $1000$~\GeV & $300$~\GeV & $1$ &$> 400$~\GeV & $6.4$ & $2.7$ \\
 & $1000$~\GeV & $300$~\GeV & $5$ &$> 400$~\GeV & $0.4$ & $3.9$ \\
 & $1200$~\GeV & $400$~\GeV & $1$ &$> 400$~\GeV & $3.3$ & $2.0$ \\ 
 & $1400$~\GeV & $300$~\GeV & $1$ &$> 400$~\GeV & $2.2$ & $0.4$ \\
 \hline \hline
    \end{tabular}
  \end{center}
\end{table*}

The boosted channel differs from the resolved channel primarily by the requirement of at least one large-$R$ jet designed to contain the decay products of a single \hbb decay.
Table~\ref{tab:eventselection} also lists the selection criteria for the boosted channel, designed to achieve high efficiency for the EFT models and good background rejection. 
The leading large-$R$ jet is required to have $\pt > 350$~\GeV. 
At these high \pt values, the decay products from top quarks are often contained inside a large-$R$ jet, so the requirement on the mass of the leading large-$R$ jet to between $90$~\GeV~and $150$~\GeV~provides good rejection against top quark background. 
The multijet background is further suppressed by requiring the azimuthal angle between $\vecMET$ and $\vecMPT$, $\Delta\phi(\vecMET,\vecMPT)$, to be less than $\pi/2$. 
Similar to the resolved channel, the final \MET requirement in the boosted channel varies as the \MET distribution shifts for different EFT models and DM mass.
For the models $|\chi|^2|H|^2$, $\bar{\chi}i\gamma_5\chi|H|^2$, and $\chi^\dagger \partial^\mu \chi  H^\dagger D_\mu H$, the minimum \MET is $300$~\GeV~for $m_\chi$ = $1$, $65$, and $100$~\GeV, and $400$~\GeV~for $m_\chi$ = $500$ and $1000$~\GeV; the selection efficiency for these three EFT models varies from $1\%$ to $8\%$, with a higher efficiency at larger $m_\chi$. 
For the $\bar \chi \gamma^\mu \chi  B_{\mu\nu} H^\dagger D^\nu H$ model, $\MET > 400$~\GeV~is required for all $m_\chi$ values, and the selection efficiency ranges from $10\%$ to $13\%$, increasing slightly with $m_\chi$.

\section{Background estimation}
\label{sec:background}
The main source of irreducible background for this search is $Z$+jets when the $Z$ boson decays into a pair of neutrinos. To reduce the impact of theoretical and experimental uncertainties associated with this process,  which are particularly evident in regions with large \MET, $Z(\rightarrow\nu\bar{\nu})$+jets background is determined from data with input from simulation, as described in Section \ref{sec:znnbackground}.  Multijet production in which there is large \MET is not simulated reliably, so it is also estimated using data, as described in Section \ref{sec:qcdbackground}. The $W(\to\ell\nu)$+jets and top quark production processes are estimated using the shape from MC simulation and are normalized to data in 1-lepton control regions, as described in Section~\ref{sec:1LepCR}.
The other backgrounds are estimated from Monte Carlo simulation, namely $Z(\to\ell\ell)$+jets, diboson production, and vector boson associated production with the Standard Model Higgs boson.
Section~\ref{sec:0LepVR} shows validations of the background modeling in the zero-lepton validation regions using selections close to those of the signal regions.

\subsection{$Z(\rightarrow\nu\bar{\nu})$+jets background}
\label{sec:znnbackground}
The estimation of the $Z(\rightarrow\nu\bar{\nu})$+jets background is derived from two data samples.  For $\Etmiss < 200$~\GeV, the $Z(\rightarrow\mu^+\mu^-)$+jets sample is used.  The \pT spectrum of produced $Z$ bosons and the kinematic distributions of jets are the same whether the $Z$ boson decays into charged leptons or neutrinos.  Thus the $Z(\rightarrow\mu^+\mu^-)$+jets data sample provides very good modeling of the $Z(\rightarrow\nu\bar{\nu})$+jets background. The $Z(\rightarrow\mu^+\mu^-)$+jets events are selected by requesting two isolated muons that pass the $24$~\GeV~muon trigger in the HLT and satisfy the {\it tight} selection criteria, with opposite charge and \pT above $25$~\GeV, and the invariant mass of the muon pair be between $70$~\GeV~and $110$~\GeV. The same selection is applied to both simulated samples and to the data. A transfer function is derived to account for the differences in branching ratio, trigger efficiency, and reconstruction efficiencies between $Z(\rightarrow\nu\bar{\nu})$+jets and $Z(\rightarrow\mu^+\mu^-)$+jets.  For higher purity and larger sample size, as well as reduction of systematic uncertainties, \Sherpa samples of $Z(\rightarrow\nu\bar{\nu})$+jets and $Z(\rightarrow\mu^+\mu^-)$+jets, which have the same production kinematics, are used to derive the transfer function.  The samples are fully reconstructed and the trigger and event selection criteria are applied.  The \MET in each $Z(\rightarrow\mu^+\mu^-)$+jets event is recalculated by adding the two muon transverse momentum vectors to the original \MET to create a new variable called \Etmiss$^{+\ell\ell}$.  This mimics the \MET in $Z(\rightarrow\nu\bar{\nu})$+jets events.  A transfer function is derived by fitting the ratio of the $Z(\rightarrow\nu\bar{\nu})$+jets \MET distribution divided by the $Z(\rightarrow\mu^+\mu^-)$+jets \Etmiss$^{+\ell\ell}$ distribution. Simulated events from other background processes that passed the aforementioned $Z(\rightarrow\mu^+\mu^-)$ selection are subtracted from the data to obtain a $Z(\rightarrow\mu^+\mu^-)$+jets data sample with high purity. The MC-based transfer function is applied to the $Z(\rightarrow\mu^+\mu^-)$+jets \Etmiss$^{+\ell\ell}$ distribution in this data sample to estimate the $Z(\rightarrow\nu\bar{\nu})$+jets background.
As the \Zprime-2HDM model contains the decay mode $\Zprime \to Z h$, the presence of such a signal would have a contribution to the $Z(\rightarrow\mu^+\mu^-)$+jets process as well; however, in the $\Etmiss < 200$~\GeV~region, the expected yield from the $\Zprime \to Z(\to\mu^+\mu^-) h$ process is several orders of magnitude smaller than the Standard Model $Z(\rightarrow\mu^+\mu^-)$+jets production, and thus has a negligible impact on the background estimation. 

For $\Etmiss > 200$~\GeV, the limited size of the $Z(\rightarrow\mu^+\mu^-)$+jets data sample reduces its usefulness. In this region the $\gamma$+jets data sample is used. For $\gamma$ (or in this case the modified \MET as described below) transverse momenta much greater than the mass of the $Z$ boson, the kinematic properties of $\gamma$+jets and $Z$+jets events are very similar~\cite{Ask:2011xf}. A high-purity (above $99\%$ in both the resolved and boosted channels after $b$-tagging requirements) $\gamma$+jets data sample is selected by requiring one high-\pT ($\geq 125$~\GeV), prompt photon that passed the $120$~\GeV~HLT photon trigger. The transfer function is calculated from reconstructed \Sherpa samples of $\gamma$+jets events that passed the same photon selection, and $Z(\rightarrow\nu\bar{\nu})$+jets events. The \MET in a $\gamma$+jets event is recalculated by using all clustered objects described in Section~\ref{sec:object} except the leading photon, and denoted as \Etmiss$^{+\gamma}$. The $Z(\rightarrow\nu\bar{\nu})$+jets background for $\Etmiss > 200$~\GeV~is obtained by multiplying the $\gamma$+jets \Etmiss$^{+\gamma}$ distribution in the data by the MC-produced transfer function. Since the photon couples to a quark through its electric charge, while the $Z$ boson coupling depends on the weak neutral vector and axial-vector couplings, 
the transfer function varies slightly by $\sim3\%$ to $10\%$ depending on the number of $b$-tagged jets in the final state. A MC-based correction factor for each value of $b$-tagged jet multiplicity is derived and applied to account for the small difference. 

\begin{figure*}[!ht]
\begin{center}
\subfloat[ Resolved channel ]{\includegraphics[width=0.495\textwidth]{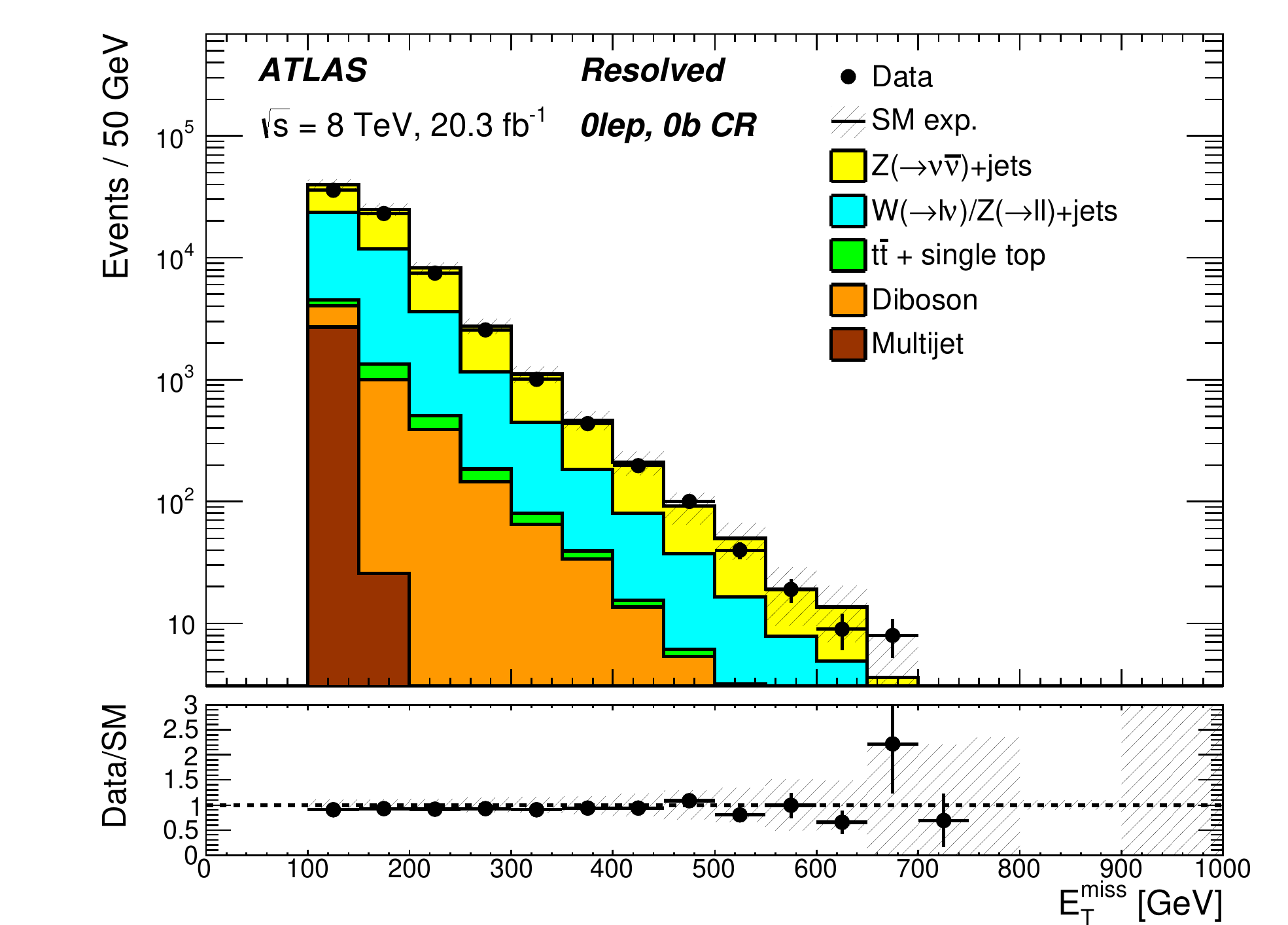}}
\subfloat[ Boosted channel ]{\includegraphics[width=0.495\textwidth]{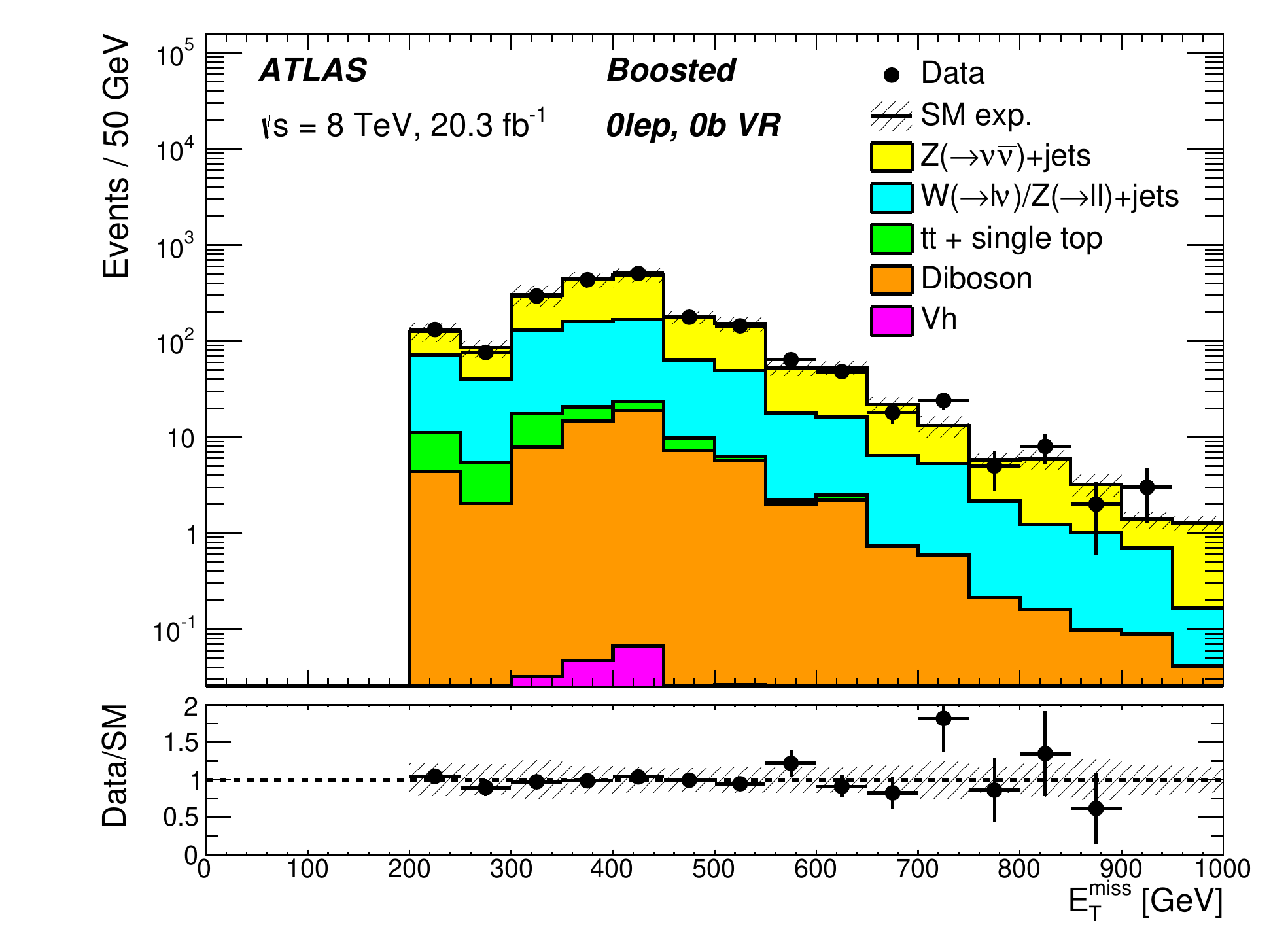}}
\end{center}
\caption{The distribution of the missing transverse momentum with magnitude \MET of (a) the resolved channel and (b) the boosted channel in the zero-lepton, zero-$b$-tagged jet control region (CR) for the estimated backgrounds (solid histograms) and the observed data (points). The hatched areas represent the combined statistical and systematic uncertainties in the total background estimation. The minimum \MET requirement in the resolved (boosted) channel is $100$~\GeV~($200$~\GeV). In the resolved channel, the small contributions from $Wh$ and $Zh$ are included in the $W$ or $Z(\rightarrow\nu\bar{\nu})$ plus jets distributions.}
\label{fig:znn0bCR}
\end{figure*}

To test this procedure over the entire \MET distribution above 100~\GeV, two control regions are defined in the resolved channel using event selection very similar to that of the signal region except requiring either zero or one $b$-tagged small-$R$ jet. A similar test is performed in the boosted channel but with $\MET$ above 200~\GeV\ where control regions are defined with zero, one or two $b$-tagged track-jets that are matched by ghost association to the leading large-$R$ jet. Despite the two $b$-tagged track-jets requirement in the last case, the expected discovery significance of the signal models considered is well below $2\sigma$ considering the background estimate. By keeping the yields of the other background processes constant and normalizing the total expected background to the data, a scale factor of $0.9$ for the $Z(\rightarrow\nu\bar{\nu})$+jets estimation is derived from the control regions with no $b$-tagged jets for both the resolved and boosted channels. The $10\%$ difference from unity is assigned as an additional source of systematic uncertainty on the $Z(\rightarrow\nu\bar{\nu})$+jets normalization in both channels. 
After the corrections described above are applied, the data and the estimated background agree well in all five control regions to within $3\%$ to $10\%$ in the resolved channel, and within $1\%$ to $20\%$ in the boosted channel; the differences are larger in regions with higher $b$-tagged jet multiplicity and hence smaller event sample size. Figure~\ref{fig:znn0bCR} shows the \MET distributions in the zero-lepton, zero-$b$-tagged jet control regions of the resolved and boosted channels. Good agreement is demonstrated between the data and the estimated background. 

\subsection{Multijet background}
\label{sec:qcdbackground}
The multijet background in the resolved channel is estimated from data using a ``jet smearing'' method~\cite{Squark:prd}. 
A pure multijet sample, used as the ``seed'' events, is obtained by selecting from the data events containing multiple jets, no isolated leptons, and \MET below $120$~\GeV, using a set of jet triggers with different requirements on jet \pT~threshold and $|\eta|$~coverage.  
A ``smeared'' event is generated by multiplying each jet four-momentum in a seed event by a random number drawn from a jet response function. The response function quantifies the probability of fluctuations in the detector response to jets measured in the data. It is determined using data and simulation, and has both Gaussian and non-Gaussian components to account for both the core of the distribution and the tails. 
After ``smearing'', the obtained multijet estimation is compared to the data in a dedicated multijet control region in which $100 < \Etmiss < 120$~\GeV, the leading jet has $\pT > 100$~\GeV, and $\Delta\phi_{\textrm{min}}(\vecMET, j_i) < 0.7$. 
The agreement is good with slight mismodelling likely due to the difference in \Etmiss distributions between $b$-quark jets and light jets. 
Hence the ``smeared'' multijet sample is reweighted two-dimensionally with respect to its jet multiplicity and $b$-tagged jet multiplicity to match the numbers in the data in the multijet control region. 
The aforementioned small discrepancies in the data and background comparison are removed after reweighting. 
The multijet background is small in the other control regions in the resolved channel and negligible in the signal region.

The multijet background is estimated in the boosted channel using an ``ABCD method''~\cite{Aad:2014yka},
in which the data are divided into four regions based on the
$\Delta\phi_{\textrm{min}}(\vecMET, j_i)$ and
$\Delta\phi(\vecMET, \vecMPT)$ variables, such that three of the regions are dominated
by the background.
These two variables are found to be weakly correlated in a data sample after the lepton veto, and requiring at least one large-$R$ jet with $\pt^{J}>350$~\GeV, at least two track-jets matched to the large-$R$ jet, and $\MET$ between $100$ and $200$~\GeV. This observation is confirmed in a multijet event sample simulated with \pythiaeight.
The signal region (A) is selected with $\Delta \phi_{\textrm{min}}(\vecMET, j_i) > 1.0$ and $\Delta\phi(\vecMET, \vecMPT) < \pi/2$.
In region C, the requirement on $\Delta\phi(\vecMET, \vecMPT)$ is reversed.
In regions B and D, $\Delta \phi_{\textrm{min}}(\vecMET, j_i)  < 0.4$ is required, with the same requirement on $\Delta\phi(\vecMET, \vecMPT) $ as in regions A and C, respectively.
The multijet yield in each of the regions B, C, and D is obtained by subtracting from the data the contribution of other backgrounds taken from simulation. 
The number of multijet events in region A is estimated as a product of the yields in regions D and C divided by the yield in region B. 
Due to the small number of events, the track-jet $b$-tagging and the large-$R$ jet mass requirements for the signal region are not applied in regions B, C, and D, and an additional scale factor to estimate the selection efficiencies of these two requirements is applied to the resulting yields.  The number of events from multijet background in the signal region is estimated to be consistent with zero within uncertainties, and a $68\%$ CL upper limit of $0.1$ events is used as the predicted yield.

\subsection{$W$+jets and top quark backgrounds}
\label{sec:1LepCR}

In the resolved channel, the $W$+jets control region is very similar to the signal region, except that the lepton veto is replaced by the requirement of one isolated muon with $\pt > 25$~\GeV, and the number of small-$R$ jets must be two. 
The purity of the $W$+jets background in this control region is approximately $90\%$ before $b$-tagging requirements. By keeping the yields of the other background processes constant and normalizing the total expected background to data, a scale factor of $0.92$ is derived for the $W$+jets background. The $8\%$ difference from unity is small compared to the systematic uncertainty on the $W$+jets normalization as discussed in Section~\ref{sec:ResolvedSyst}. This scale factor is applied to the $W$+jets background when deriving the normalization for $Z(\rightarrow\nu\bar{\nu})$+jets background in Section \ref{sec:znnbackground}. 
The top quark control region has the same requirements except that three small-$R$ jets are required. The purity of the top quark background, which includes mostly \ttbar but also single-top-quark events, is approximately $78\%$  in the top quark control region after requiring at least one $b$-tagged small-$R$ jet. Good agreement is observed between the data and simulation and no additional scale factor is applied to the top quark background. 
In both control regions, as well as the combined one-lepton validation region where the jet multiplicity requirement is removed, there is good agreement between the data and estimated background in both number of events and modeling of the kinematic variables. 

As Monte Carlo simulation predicts a larger fraction of high \pt top quarks in \ttbar events than is seen in the data, a correction is applied in the boosted channel at the level of generated top quarks in the \ttbar MC sample~\cite{Aad:2014zka, Aad:2015gra}. 
For the resolved channel, the correction is not applied since the impact is small, but the effect of it is accounted for as a source of systematic uncertainty, as discussed in Section~\ref{sec:ResolvedSyst}. 

The $W$+jets and top quark (\ttbar+ single top quark) backgrounds are further studied in the boosted channel in a one-lepton control region selected by requiring one isolated muon with $\pt > 25$~\GeV, preselection criteria as described in Section~\ref{tab:eventselection} except the lepton veto, and the first two selections in Table~\ref{tab:eventselection}.
Events passing the one-lepton control region selections are categorized as being in the $W$+jets control region unless at least one $b$-tagged track-jet is found within $\DeltaR=1.5$ of the muon direction, in which case they are used for a top quark control region. The purity of $W$+jets background in the $W$+jets control region is approximately 72\%, whereas the purity of the top quark background in the top quark control region is $\sim90\%$. A pair of linear equations to calculate the normalization factor from background to the data is constructed using the predicted and observed yields of the $W$+jets and top quark backgrounds. The solution of the equations, $0.82 \pm 0.05$ and $0.89 \pm 0.06$, are applied as scale factors to the $W$+jets background and top quark background, respectively.

\subsection{Zero-lepton validation region}
\label{sec:0LepVR}

\begin{figure*}[!ht]
\begin{center}
\subfloat[Resolved channel, \MET]{\includegraphics[width=0.495\textwidth]{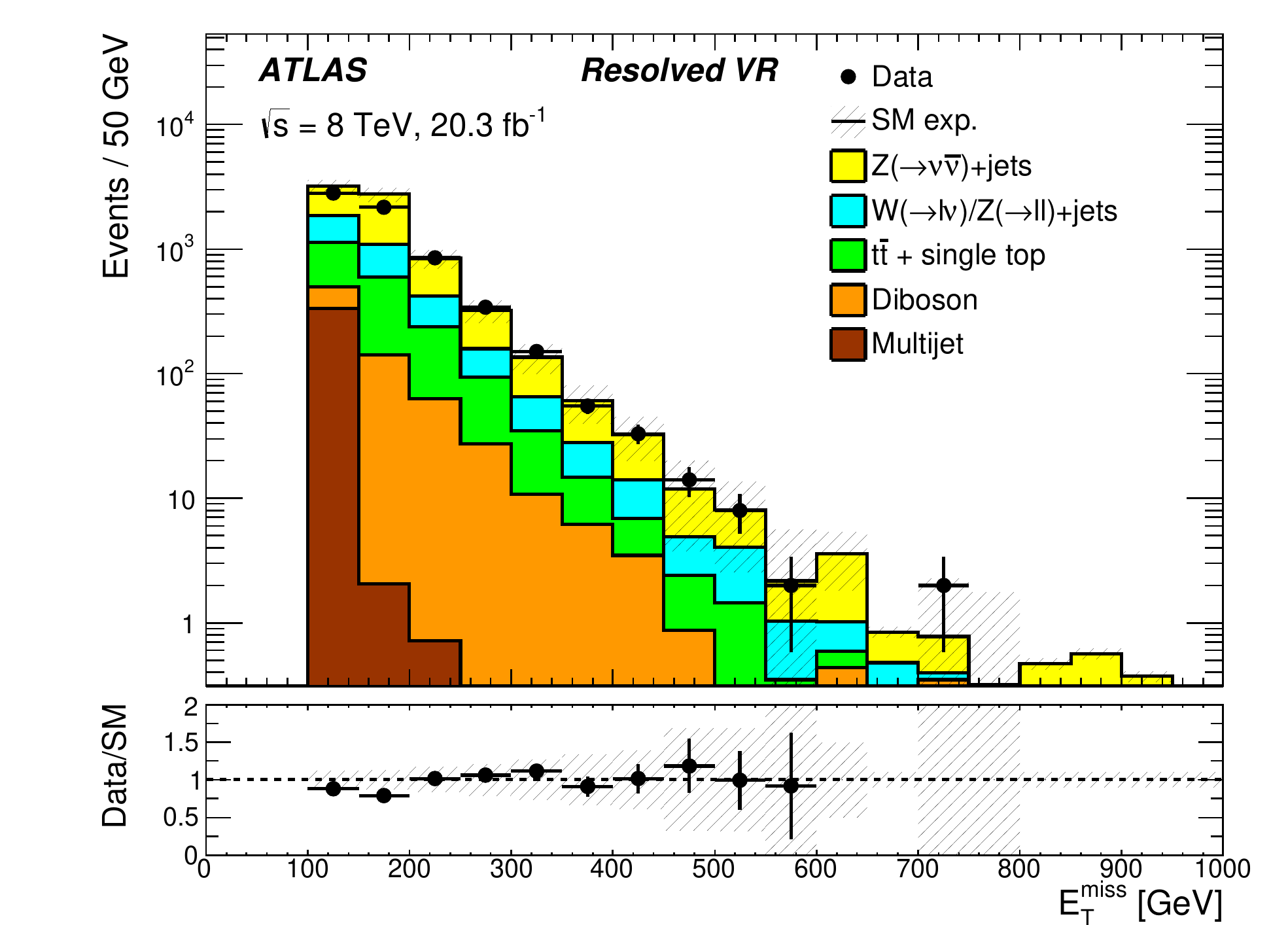}\label{fig:metR}}
\subfloat[Boosted channel, \MET]{\includegraphics[width=0.495\textwidth]{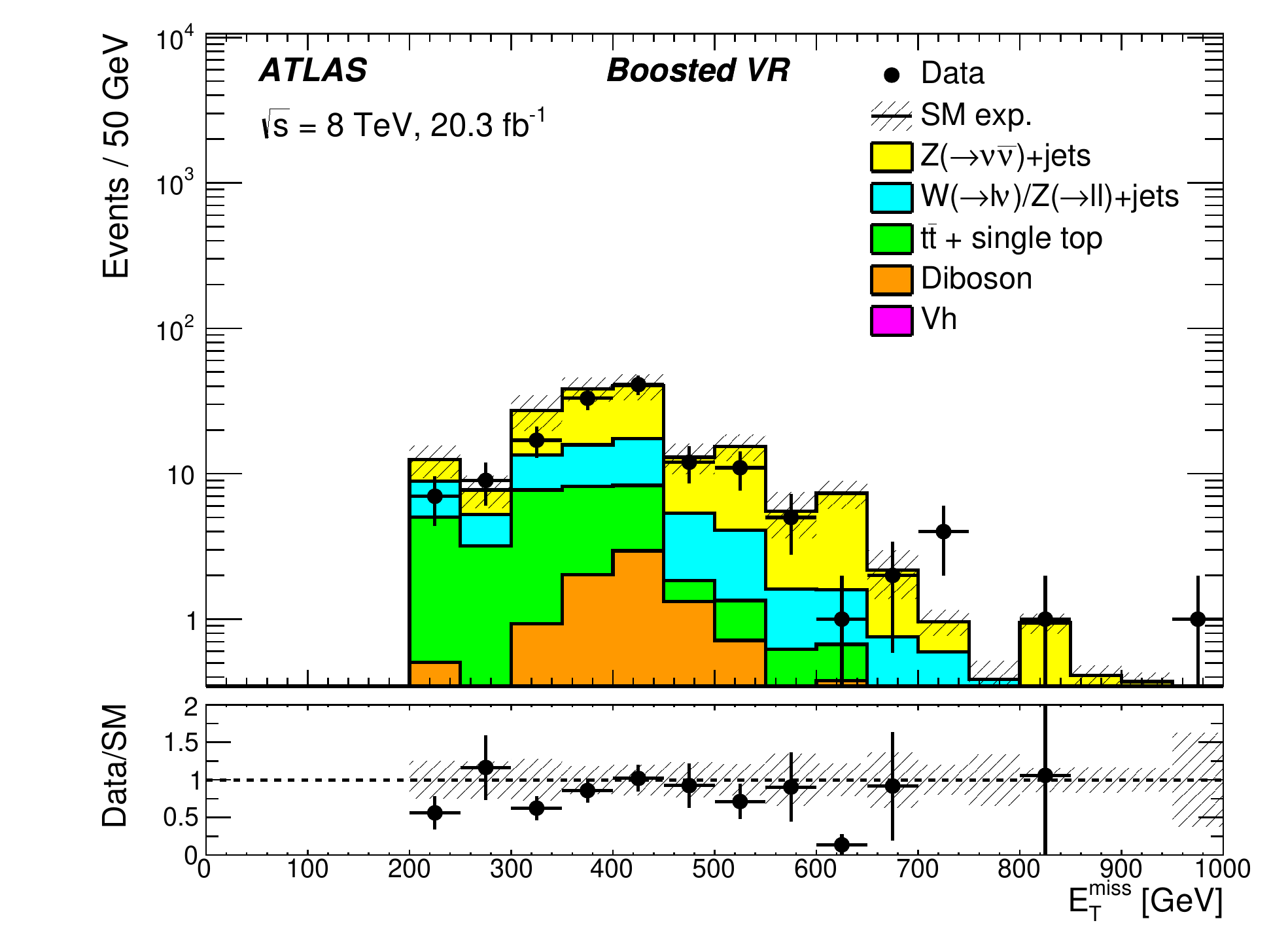}\label{fig:metB}}
\\
\subfloat[Resolved channel, $m_{b_{1}b_{2}}$ ]{\includegraphics[width=0.495\textwidth]{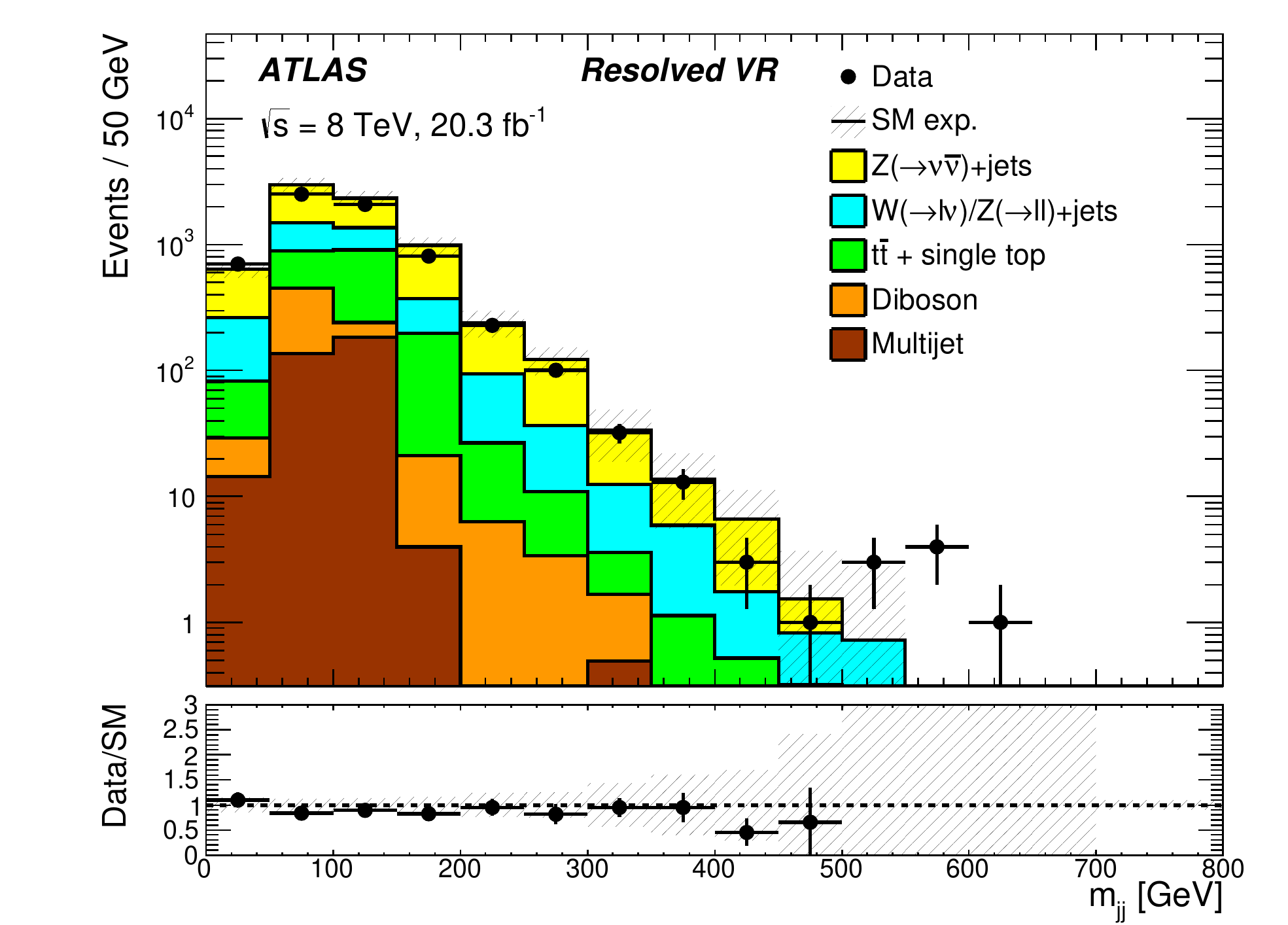}\label{fig:mbbR}}
\subfloat[Boosted channel,  $m_{J_{1}}$ ]{\includegraphics[width=0.495\textwidth]{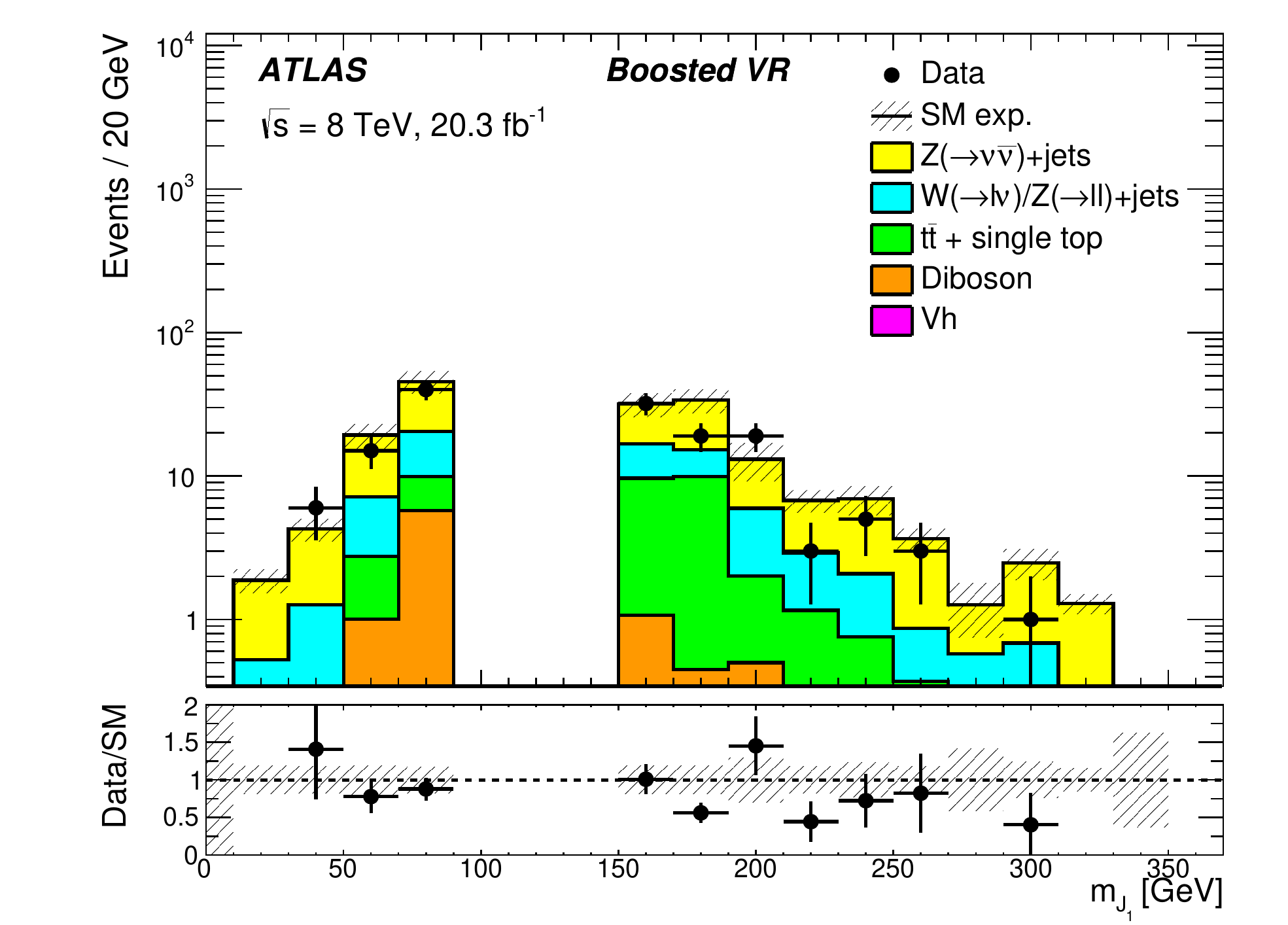}\label{fig:mbbB}}
\end{center}
\caption{Distributions of the missing transverse momentum with magnitude \MET for (a) the resolved channel and (b) the boosted channel and the invariant mass distributions for (c) the two leading small-$R$ jets in the resolved channel and (d) the leading large-$R$ jet in the boosted channel. Events are selected in the zero-lepton validation region (VR) for the estimated backgrounds (solid histograms) and the observed data (points). The hatched areas represent the combined statistical and systematic uncertainties in the total background estimation. At least one (exactly one) $b$-tagged jet is required in the resolved (boosted) channel. In the resolved channel, the invariant mass of the \bbbar system in events with at least two $b$-tagged jets is required to be either less than $60$~\GeV~or greater than 150~\GeV. In the boosted channel, the invariant mass of the large-$R$ jet with exactly one $b$-tagged track-jet is required to be either less than $90$~\GeV~or greater than 150~\GeV. The minimum \MET requirement in the resolved (boosted) channel is $100$~\GeV~($200$~\GeV). In the resolved channel, the small contributions from $Wh$ and $Zh$ are included in the $W$ or $Z(\rightarrow\nu\bar{\nu})$ plus jets distributions.}
\label{fig:VRMet2}
\end{figure*}

The individual background processes are studied and normalized to the data in the dedicated control regions, as described in the previous sections. To examine the overall modeling of all non-Higgs background processes combined, zero-lepton validation regions are defined for both channels, with selections similar to the signal region, but reversing the requirement on the invariant mass of the \bbbar system.
In the resolved channel, events are selected with at least one $b$-tagged small-$R$ jet, and for events with two or more $b$-tagged jets, the invariant mass of the two leading $b$-tagged jets is required to be either below $60$~\GeV~or above $150$~\GeV. 
In the boosted channel, events are selected with exactly one $b$-tagged track-jet associated with the leading large-$R$ jet, and the invariant mass of the large-$R$ jet is required to be either below $90$~\GeV~or above $150$~\GeV. 
Figure~\ref{fig:VRMet2}\subref{fig:metR} and Figure~\ref{fig:VRMet2}\subref{fig:metB} show the \MET~distributions in both channels, and Figure~\ref{fig:VRMet2}\subref{fig:mbbR} and Figure~\ref{fig:VRMet2}\subref{fig:mbbB} show the distribution of the invariant mass of the two leading small-$R$ jets (the invariant mass of the leading large-$R$ jet) in the resolved (boosted) channel. 
The aforementioned scale factors for the corresponding background processes have been applied. Good agreement between the data and the estimated background is achieved for different kinematic variables, including jet \pT, angular distributions, multiplicity, and number of $b$-tagged jets, at each selection stage in both channels.  

\begin{figure*}[!ht]
\begin{center}
\subfloat[Resolved channel]{\includegraphics[width=0.495\textwidth]{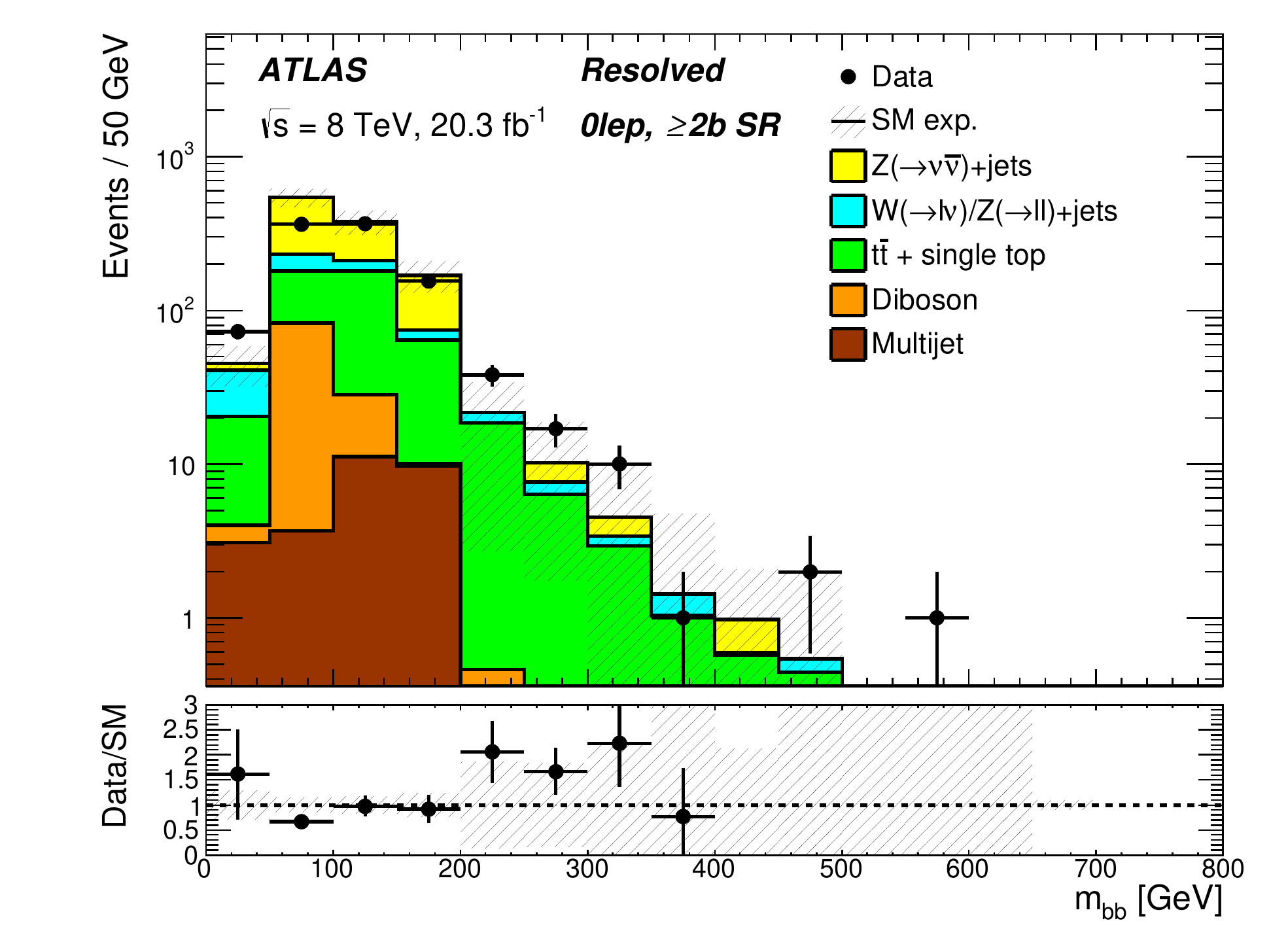}}
\subfloat[Boosted channel]{\includegraphics[width=0.495\textwidth]{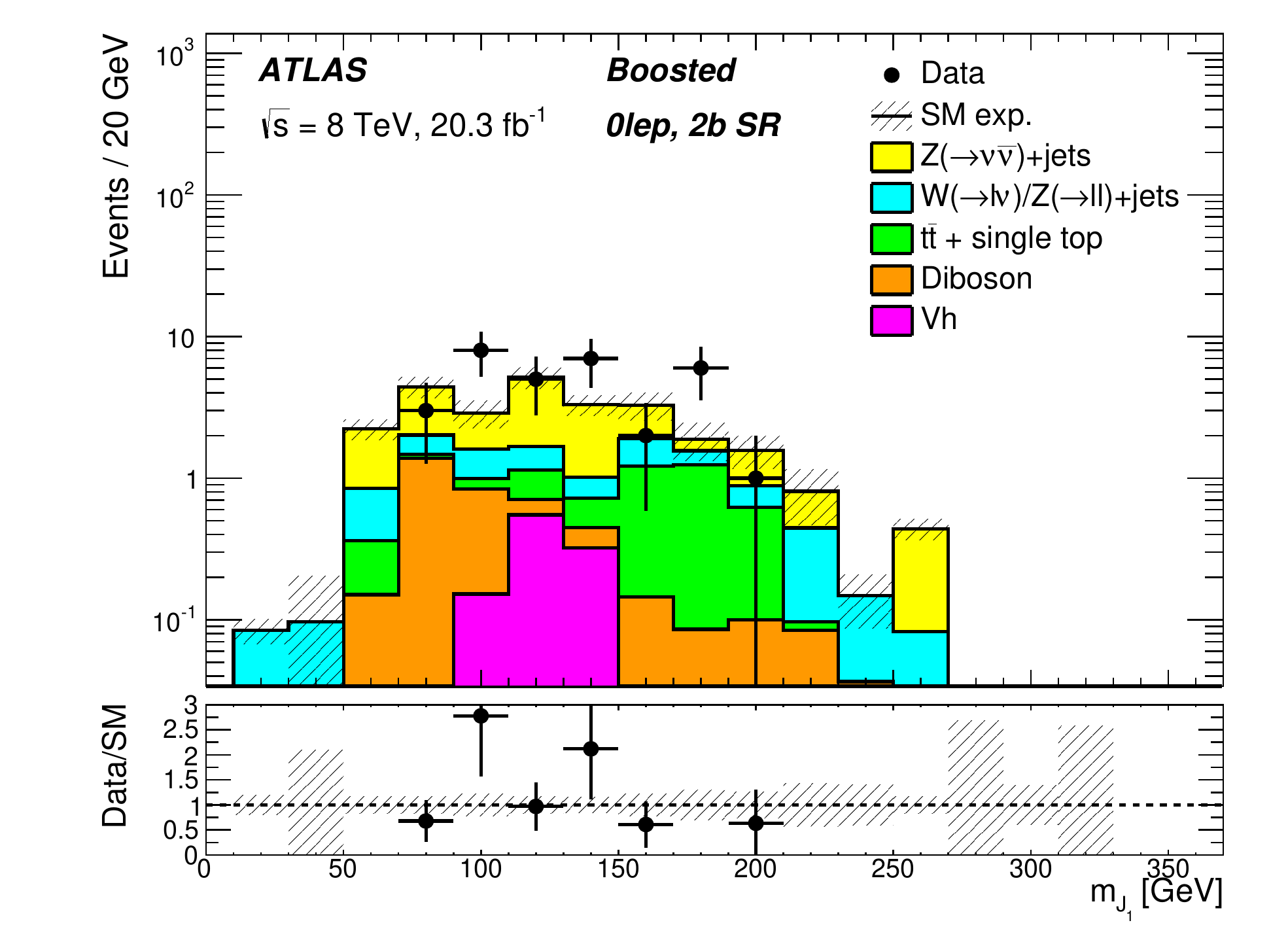}}
\end{center}
\caption{The distributions of the invariant mass of the \bbbar system for the estimated backgrounds (solid histograms) and the observed data (points) in (a) the resolved and (b) the boosted channels in the signal region (SR) without the requirement on the invariant mass. The regions with the invariant mass of the \bbbar system  between $90$~\GeV~and $150$~\GeV~are the signal regions for both channels. The hatched areas represent the combined statistical and systematic uncertainties in the total background estimation. The minimum \MET is required to be $100$~\GeV~($200$~\GeV) in the resolved (boosted) channel. At least (exactly) two $b$-tagged small-$R$ jets (track-jets) are required in the resolved (boosted) channel. In the resolved channel, the small contributions from $Wh$ and $Zh$ are included in the $W$ or $Z(\rightarrow\nu\bar{\nu})$ plus jets distributions.}
\label{fig:VRmbb}
\end{figure*}

Figure~\ref{fig:VRmbb} shows the distributions of the invariant mass of the \bbbar system in both the resolved and boosted channels with fully hadronic selection very similar to the signal region, but removing the requirement on the invariant mass. 
The regions with the invariant mass of the \bbbar system  between $90$~\GeV~and $150$~\GeV~are the signal regions for both channels. 
The signal regions were blinded in this analysis until all the studies in the aforementioned control regions and validation regions were complete. 

\section{Systematic uncertainties}
\label{sec:ResolvedSyst}
The systematic uncertainty on background estimation and signal processes using Monte Carlo samples comes from several sources, and is evaluated for each of the signal and background processes in both channels. The uncertainty associated with the $b$-tagging efficiency, which is determined from comparisons between simulation and heavy-flavor-enriched data samples~\cite{Aad:2015ydr}, ranges from $\sim 10\%$ to 15\%. 
The uncertainty on the overall background estimate due to light-flavor and charm-quark jets being misidentified as $b$-quark jets is calculated to be $\sim 1\%$ for small-$R$ jets, and $\sim 2\%$ to $3\%$ for track-jets. 
The jet energy scale and resolution~\cite{Aad:2014bia}, which directly impact the \Etmiss, depend on the kinematic properties of the jet, the distance to its nearest jet neighbor, and the flavor of the initiating parton.  The systematic uncertainty associated with the jet energy scale and resolution ranges from $\sim 5\%$ to 15\%. 

In the boosted channel, the invariant mass of the \bbbar system from the Higgs boson decay is selected by requiring the mass of the large-$R$ jet to be between $90$~\GeV~and $150$~\GeV, leading to additional systematic uncertainties from the jet mass scale and resolution~\cite{Aad:2013gja}. The uncertainties associated with jet mass are $\sim 1\%$ for the EFT signals and $\sim 3\%$ to 8\% for most simulated background processes. 
While the large-$R$ jet calibration and uncertainty are derived primarily using an inclusive multijet sample, the large-$R$ jet selection in this analysis focuses specifically on identifying jets containing two $b$-hadrons. As such, there are possible additional sources of uncertainty on the modeling of the jet mass and energy due to the difference in heavy-flavor content between the calibration and analysis selections. However, studies of multijet samples enriched with jets containing two $b$-hadrons suggest that this uncertainty is small in comparison to the existing uncertainty on jet mass and energy, and thus no additional uncertainty is applied. 

The uncertainty on \MET originating from the energy scale and resolution of energy clusters not included in jets~\cite{etmissref} is small at $\sim 1\%$ or less, as are the uncertainties due to possible mismodeling of the effect of multiple $pp$ collisions (pileup) and the method of removing jets coming from pileup. The uncertainty on the integrated luminosity for the data sample is 2.8\%. It is derived using the same methodology as that detailed in Ref.~\cite{Aad:2013ucp}. 

The cross-section uncertainties for the background processes are as follows.  For $t\bar{t}$ production, an uncertainty of $7\%$ is cited from theoretical calculations~\cite{Moch:2008ai}, which is consistent with the ATLAS measurement of top quark pair production~\cite{Aad:2015pga}. The same uncertainty is used for the small single-top-quark background~\cite{Kidonakis:2010ux}.  For $W$+jets, a cross-section uncertainty of 20\% is taken from the recent ATLAS measurement of $W$+jets production with $b$-jets~\cite{Aad:2013jhep}.  
The uncertainty on the simulated diboson background cross-section increases with the \MET threshold from $20\%$  for $\MET > 150$~\GeV~to $30\%$ for $\MET > 400$~\GeV~\cite{Aad:2015zva}.
For vector boson plus Higgs boson production, an uncertainty of 3.1\% on the cross-section is estimated from theoretical calculations~\cite{LHC:crosssection} and is applied here.
The signals samples from MC simulation are produced at LO. An estimated value of $10\%$ is used as the uncertainty on the signal cross-section from NLO corrections~\cite{Haisch:2013ata}. 
The systematic uncertainty on the signal acceptance due to the choice of parton distribution functions (PDFs) is determined by using the uncertainty eigenvectors provided for multiple PDF sets per the PDF4LHC prescription~\cite{Botje:2011sn}. 
The uncertainty from this source is given by the maximum difference in detector acceptance of the signal process when using different variations in the MSTW2008 LO~\cite{Martin:2009nx} and NNPDF2.1~\cite{Ball:2012cx} PDF sets, 
leading to an uncertainty of $\sim 4\%$ to $8\%$ for the \Zprime-2HDM model, and $\sim 2\%$ to $21\%$ for the different EFT models.
For the simulated background processes, the uncertainty due to variations in MSTW2008 NNLO~\cite{Martin:2009nx, Martin:2009bu}, CT10 NNLO~\cite{CT10, Gao:2013xoa}, and NNPDF2.3~\cite{Ball:2012cx} PDF sets and parton shower models is $\sim 5\%$ to $7\%$. 

The systematic uncertainty on the data-driven $Z(\rightarrow\nu\bar{\nu})$+jets background comes from the transfer function and from the simulated backgrounds that are subtracted from the $Z(\rightarrow\mu^+\mu^-)$+jets data sample (the high-\pT $\gamma$+jets sample has a purity of over $99\%$ after $b$-tagging requirements).  For the latter, all of the systematic uncertainties noted above are calculated for simulated samples.  Since these backgrounds are subtracted here, the uncertainties are anticorrelated with the variations of the corresponding backgrounds in the signal region.  For the transfer function, there are contributions from the functional form used, the stage of event selections from which the transfer function is calculated, the fit range in \Etmiss, how well the transfer function describes the shape of the ratio distribution, and the statistical uncertainty on the fit function parameters. In the high-\MET region where $\gamma$+jets simulation is used to derive the transfer function, there are additional sources of systematic uncertainty on the transfer function from the efficiencies of photon identification, reconstruction, and isolation, and photon energy scale and resolution~\cite{Aad:2014nim}. 
A $10\%$ uncertainty on the cross-section is also taken into account from the normalization factor of $0.9$ applied to the $Z(\rightarrow\nu\bar{\nu})$+jets background, as described in Section \ref{sec:znnbackground}. The theoretical uncertainty on the $Z/ \gamma$~ratio at high \pT is $\sim 4\%$~\cite{Ask:2011xf}, which is small in comparison and hence not applied.  
The total systematic uncertainty on the $Z(\rightarrow\nu\bar{\nu})$+jets background in the resolved channel is 20\% in the lower \MET region where $Z(\rightarrow\mu^+\mu^-)$+jets is used and 12\% in the higher \MET region where $\gamma$+jets is used. In the boosted channel, only $\gamma$+jets is used to estimate $Z(\rightarrow\nu\bar{\nu})$+jets background and the total systematic uncertainty is approximately 16\%.

As explained in Section \ref{sec:1LepCR}, the top quark $\pt$ distribution is reweighted at the Monte Carlo generator level to bring it into agreement with measurements of the data. 
The size of the correction is found to be $5.5\%$ in shape and normalization combined in the resolved channel, where it is considered as an additional source of systematic uncertainty. The correction has a greater effect in the boosted channel as the original mismodeling in simulation is primarily in high-\pT regions. The systematic uncertainty associated with the top quark \pT reweighting is evaluated to be $\sim 15\%$ and applied to the top quark process in the boosted channel. 

Overall, the systematic uncertainty on the estimated background is calculated to be between $10\%$ and $16\%$ in the resolved channel, and between $12\%$ and $14\%$ in the boosted channel, depending on the final \MET requirement in the signal region. Table \ref{tab:systematics_comb} lists the main sources of systematic uncertainty for both the resolved and boosted channels, and their values for both signals and backgrounds. 
The values given for the backgrounds are the uncertainties on the total background with the relative weights and correlations of individual background processes taken into account.

\begin{table*}[h]
  \begin{center}
   \caption{Summary of systematic uncertainty in percent for all backgrounds combined and signal samples in the resolved and boosted channels. The first column lists the main sources of systematic uncertainty, where the acronym JES refers to the jet energy scale, JER the jet energy resolution, JMS the jet mass scale, JMR the jet mass resolution, and JVF the jet vertex fraction. The uncertainty figures listed for ``$b$-tagging'' combine the uncertainty from both $b$-tagging efficiency and mistag rates. The uncertainty ranges in ``Total Background'' reflect the shift in value with increasing \MET threshold in the final signal region. The uncertainties for ``$Z(\nu\bar{\nu})$ transfer function'' take into account the fractional weight of the $Z(\nu\bar{\nu})$ process in total background, which differs per analysis channel and \MET threshold. Most of the systematic uncertainties on the signal models vary little across the parameter space in this analysis, with the exception of signal PDF and $\alpha_{\rm s}$, JMS, and pileup uncertainty; hence the ranges of values are shown. }
    \label{tab:systematics_comb}
    \begin{tabular}{l|cc|cc}
      \hline \hline
                                    & \multicolumn{2}{c|}{Resolved (\%)}  & \multicolumn {2}{c}{Boosted (\%)} \\
                                    & $\Zprime$-2HDM  & Total Background  & EFT & Total Background \\ \hline \hline
      $b$-tagging               	& 14 		& 6--10 	& 13		& 5.3 \\ 
      JES(small+large-$R$)   & 2.4  	& 1.8--2.8 	& 3.0 	& 2.2--8.5 \\
      JER(small+large-$R$)   & 0.6 	&  3.5--5.4 & 1.0 	& 1.5--4.6 \\
      JMS(large-$R$)             &  - 		& - 		& 1.0--2.5 	& 1.3 \\
      JMR(large-$R$)         	& - 		&  - 		& 2.0 	& 1.6 \\
      JVF  (small-$R$)         	& 0.7  	& 0.5--0.9 	& 1.1 	& 0.2--0.6\\
      $\MET$ resolution/scale       	& 0.0 	& $<0.2$  	& 0.5 	& 0.1--0.8 \\
      Pileup               		& 0.3  	& 0.1 	& 0.1--1.7 	& 2.4\\
      Cross-section          	& 10  	& 6.0--11 	& 10  	& 7.6--8.1 \\
      PDF and $\alpha_{\rm s}$      	& 3.8--7.0 	& 2.9 	& 2.0--21 	& 1.8\\
      $Z(\nu\bar{\nu})$ transfer function 		& - 	& 1.4--2.7  & -  	& 5.4--5.8 \\ \hline
     Total syst. 			& 18--19 	& 10--16 	& 13--25 	& 13--14 \\  \hline
    \end{tabular}
  \end{center}
\end{table*}

\section{Results}
\label{sec:results}

\begin{figure*}[!ht]
\begin{center}
\subfloat[Resolved channel]{\includegraphics[width=0.495\textwidth]{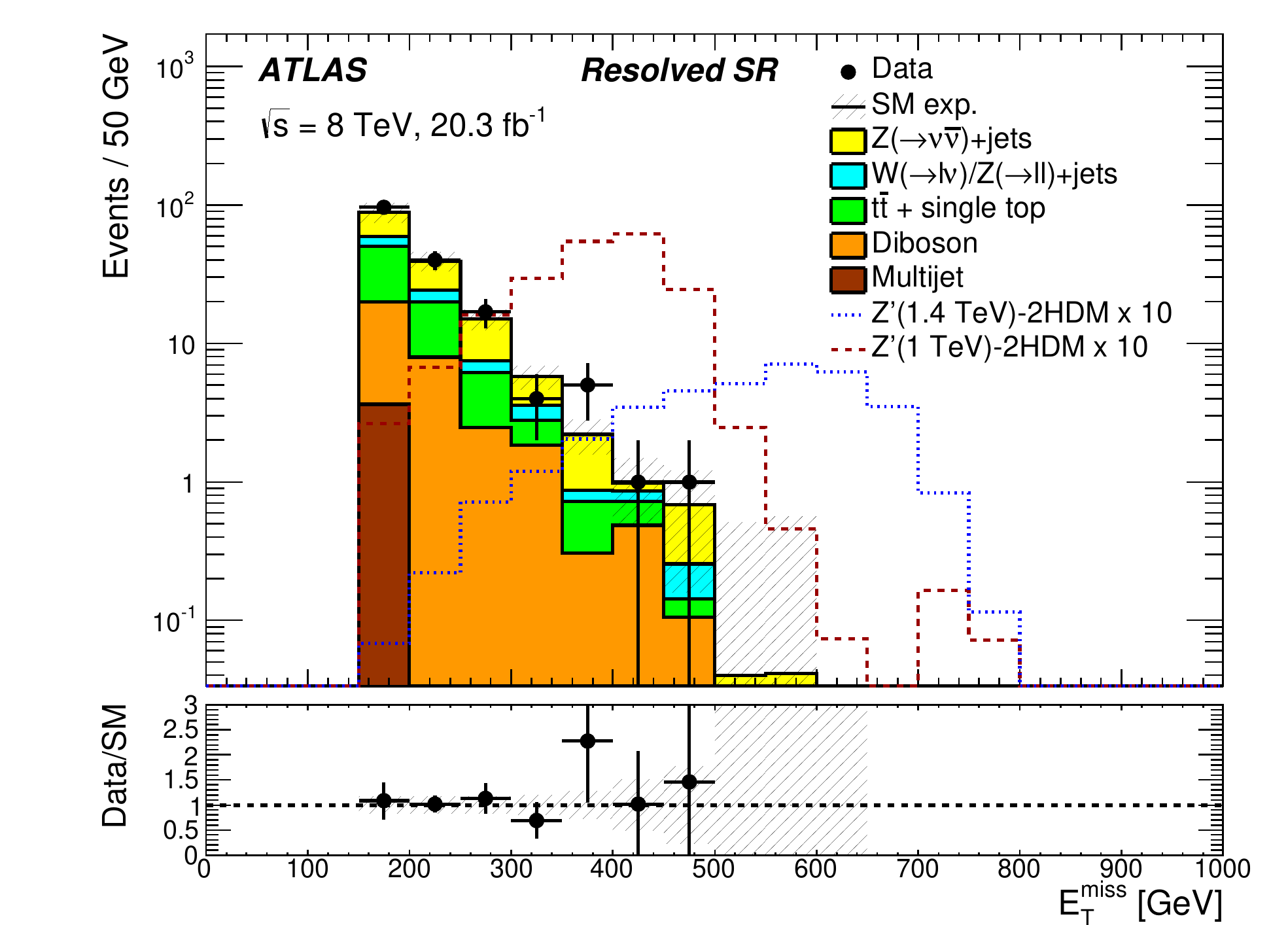}}
\subfloat[Boosted channel]{\includegraphics[width=0.495\textwidth]{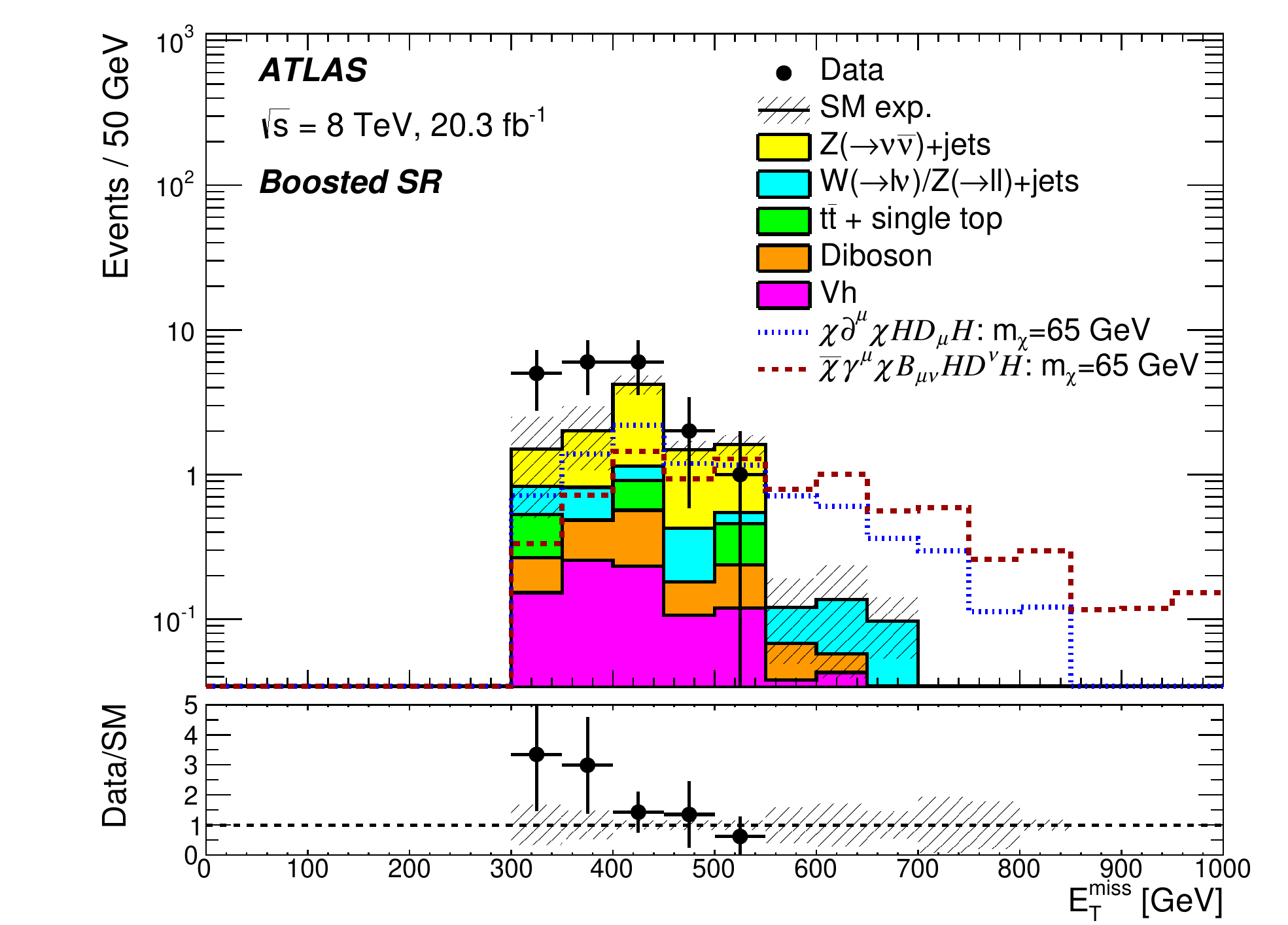}}
\end{center}
\caption{The \MET distributions of (a) the resolved channel and (b) the boosted channel in the signal region (SR) for the estimated backgrounds (solid histograms) and the observed data (points). The hatched areas represent the combined statistical and systematic uncertainties in the total background estimation. The \MET distributions for a few signal processes are overlayed in dashed lines for shape comparison: the \Zprime-2HDM signals are scaled by a factor of $10$, and the EFT signals are scaled to their corresponding expected cross-section limit. In the resolved channel, the small contributions from $Wh$ and $Zh$ are included in the $W$ or $Z(\rightarrow\nu\bar{\nu})$ plus jets distributions.}
\label{fig:resMet2}
\end{figure*}

Table~\ref{tab:resolvedRslts} shows the predicted number of background events in the signal region for each value of the ascending \MET thresholds, along with the number of events observed in the data. The numbers of predicted background events and observed events are consistent within $1\sigma$ in five out of the six signal regions. 
For the boosted channel and $\MET > 300$~\GeV, $20$ events are observed in the data compared to a background expectation of $11.2\pm2.3$ events. 
The probability that the number of events in the background fluctuates to the value in the data or above corresponds to $2.2\sigma$. 
Figure~\ref{fig:resMet2} shows the \MET distributions for the data and the estimated background in the signal regions of the resolved and boosted channels. Also shown in the resolved channel are the \MET distributions for two examples of the \Zprime-2HDM model at different $m_{\Zprime}$ with $m_{A}=300$~\GeV~and $\tan{\beta}=1$. Similarly the \MET distributions for two examples of the EFT models with different $m_{\chi}$ are shown in the boosted channel. 
The $2.2\sigma$ upward fluctuation mentioned above is primarily due to events with \MET values between $300$~\GeV~and $400$~\GeV, and mass of the leading large-$R$ jet below the Higgs boson mass, while signal events are most likely to have higher \MET values and leading large-$R$ jet mass close to Higgs boson mass. 

\begin{table*}[!ht]
\begin{center}
\caption{The numbers of predicted background events for each background process, the sum of all background components, and observed data in the signal region (SR) of the resolved and boosted channels for each of the sliding \MET requirements. Statistical and systematic uncertainties are combined. The uncertainties on the total background take into account the correlation of systematic uncertainties among different background processes. The large uncertainty on the $Z(\rightarrow\nu\bar{\nu})$+jets process in the $\MET > 150$~\GeV~SR of the resolved channel is due to limited statistics in the $Z(\rightarrow\mu^+\mu^-)$+jets data sample used for the estimation of $Z(\rightarrow\nu\bar{\nu})$+jets with $\MET < 200$~\GeV. }
\label{tab:resolvedRslts}

\begin{tabular}{l|c|c|c|c|c|c}
\hline
 & \multicolumn{4}{c|}{Resolved} & \multicolumn{2}{c}{Boosted} \\ \hline
\MET & $> 150$~\GeV & $>200$~\GeV & $>300$~\GeV & $>400$~\GeV & $> 300$~\GeV & $>400$~\GeV \\ \hline
\hline
 $Z(\rightarrow\nu\bar{\nu})$+jets 	& $48\pm32$ 	& $21\pm5$  	& $2.9\pm1.1$   & $0.3\pm0.3$ 	& $7.0\pm2.0$ 	& $5.2\pm1.6$ \\ \hline
Multijet 						& $3.7\pm3.1$  	& $0.02\pm0.02	$ & -- 		& --  			& $<0.0\pm0.1$ 	& $<0.0\pm0.1$ \\ \hline
\ttbar \& single-top 				&  $48\pm10$  	& $17\pm3.8$  	& $1.6\pm0.5$  	& $0.3\pm0.1$  	& $0.8\pm0.5$  	& $0.6\pm0.4$ \\ \hline
$W$+jets \& $Z$+jets 				& $15\pm3.4$  	&$6.2\pm1.5$ 	&$1.1\pm0.3$ 	& $0.3\pm0.1$ 	& $1.4\pm0.7$  	& $0.8\pm0.4$ \\ \hline
Diboson				 		&  $29.4\pm7.5$  	& $13.2\pm3.8$ 	& $2.8\pm1.0$ 	& $0.6\pm0.3$ 	& $0.9\pm0.5$ 	& $0.6\pm0.3$ \\ \hline
$Vh(bb)$ 						& $5.0\pm0.7$ 	& $4.2\pm0.6$ 	& $1.0\pm0.2$ 	& $0.3\pm0.1$ 	& $1.0\pm0.2$ 	& $0.6\pm0.1$ \\ \hline \hline
Total background 				&$148\pm30$	& $62\pm7.5$ 	&$9.4\pm1.8$ 	&$1.7\pm0.5$  	& $11.2\pm2.3$ 	& $7.7\pm1.7$ \\ \hline
\hline
Data 							& 164 & 68 & 11 & 2 & 20 & 9 \\ \hline
\end{tabular}
\end{center}
\end{table*}

A Frequentist approach is used for the statistical interpretation of the results~\cite{Baak:2014wma}. 
For this single bin counting experiment, the Poisson probability of the background-only hypothesis, the $p(s = 0)$-value, is calculated for each of the four signal regions with ascending \MET threshold in the resolved channel and the two signal regions in the boosted channel. 
The $95\%$ CL upper limits on the number of non-Standard Model events in each of the signal regions are also obtained using a profile-likelihood-ratio test following the $CL_s$ prescription~\cite{Read:2002hq},
which can be translated into model-independent $95\%$ CL upper limits on the visible cross-section, defined as the product of production cross-section, acceptance, and reconstruction efficiency of any signal model. 
The limits are calculated taking into account the uncertainty on the background estimate, the integrated luminosity of the data sample, and its uncertainty. 
Table~\ref{tab:limit_BSM} gives the model-independent 95\% CL upper limits on the visible cross-section, the observed and expected limits on the number of non-Standard Model events in the signal region, and the $p(s = 0)$-values.  

As a $p(s = 0)$-value of $0.03$~is calculated for \MET $> 300$~\GeV~in the boosted channel, a calculation of the look-elsewhere effect~\cite{2010EPJC...70..525G} is performed. 
Using pseudo-experiments and taking into account correlations between all signal regions in both channels, the probability that there is a deviation in the data from the background expectation at least as significant as the one observed due to a statistical fluctuation in the background is calculated to be approximately $10\%$. 

\begin{table*}[!ht]
\begin{center}
\caption{Model-independent upper limits for the resolved and boosted channels. Left to right: signal region (SR) \MET~requirement, number of observed events, number of expected background events, 95\% CL upper limits on the visible cross-section
($\langle \sigma_{\mathrm{vis}} \rangle_{\rm obs}^{95}$) and the number of non-SM events (${N_{\mathrm{BSM}}}_{\rm obs}^{95}$).  The sixth column (${N_{\mathrm{BSM}}}_{\rm exp}^{95}$) shows the expected 95\% CL upper limit on the number of non-SM events, given the estimated number and the $\pm 1\sigma$  uncertainty of background events. The last column shows the $p$-value for the background-only hypothesis ($p(s = 0)$).}
\label{tab:limit_BSM}

\begin{tabular}{llcccccc}
\hline \hline
 &  \MET & $N_{\mathrm{obs}}$ & $N_{\mathrm{bkgd}}$ & $\langle \sigma_{\mathrm{vis}} \rangle_{\rm obs}^{95}$[fb] & ${N_{\mathrm{BSM}}}_{\rm obs}^{95}$ & ${N_{\mathrm{BSM}}}_{\rm exp}^{95}$ & $p(s=0)$  \\ \hline
\multirow{4}{*}{Resolved} & $> 150$~\GeV & 164 & 148 & $3.6$ &  $74$  & ${63}^{+22}_{-14}$ &  $0.31$ \\
 & $> 200$~\GeV & 68 & 62 & $1.3$ &  $27$  & ${21}^{+8.4}_{-3.9}$ &  $0.28$ \\
 & $> 300$~\GeV & 11 & 9.4 & $0.49$ &  $9.9$  & ${8.2}^{+3.4}_{-1.9}$ &  $0.31$  \\
 & $> 400$~\GeV & 2 & 1.7   & $0.24$ &  $4.8$  & ${4.7}^{+1.6}_{-1.0}$ &  $0.39$ \\ \hline
\multirow{2}{*}{Boosted} & $> 300$~\GeV & 20 & 11.2  & 0.90 & 18 & ${9.9}^{+4.2}_{-2.9}$ & 0.03 \\
 &  $> 400$~\GeV & 9 & 7.7 & 0.43 & 8.8 & ${7.7}^{+3.3}_{-2.0}$ & 0.37 \\ \hline
\hline
\end{tabular}
\end{center}
\end{table*}

The numbers of observed events and expected background events, along with each of the signal and background statistical and systematic uncertainties, are used to determine limits for the \Zprime-2HDM model and EFT models, which are interpreted separately. Limits on the signal yield are set using a similar profile-likelihood-ratio test with the $CL_s$ method as the aforementioned model-independent upper limit calculation. Each of the systematic uncertainties is treated as a nuisance parameter, with the correlations among the sources of systematic uncertainty taken into account. 

\begin{figure}[!ht]
\begin{center}
\subfloat[$m_{\Zprime} - m_{A}$ ]{\includegraphics[width=0.5\textwidth]{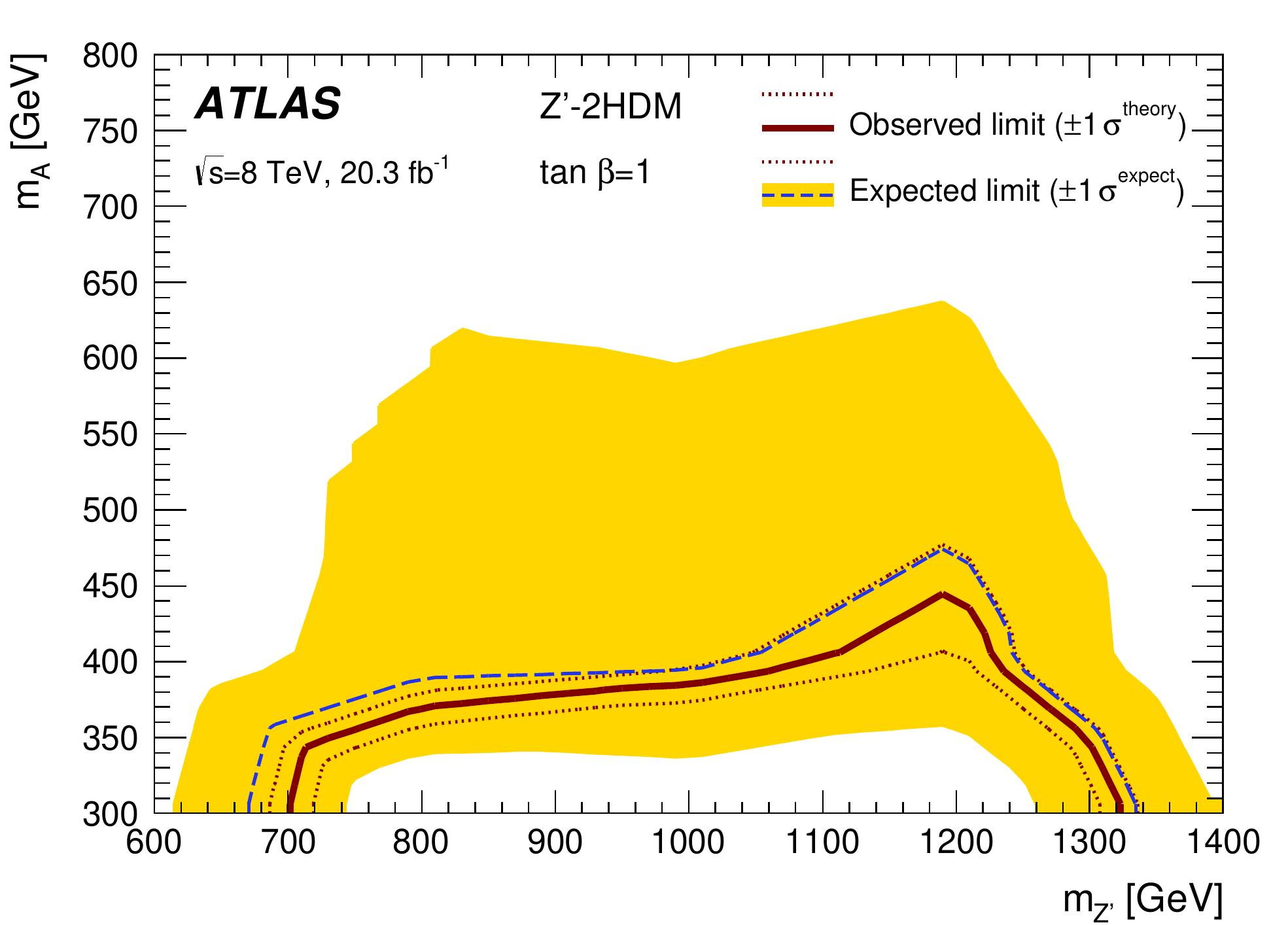}\label{fig:mA}}
\subfloat[$m_{\Zprime} - \tan{\beta}$ ]{\includegraphics[width=0.5\textwidth]{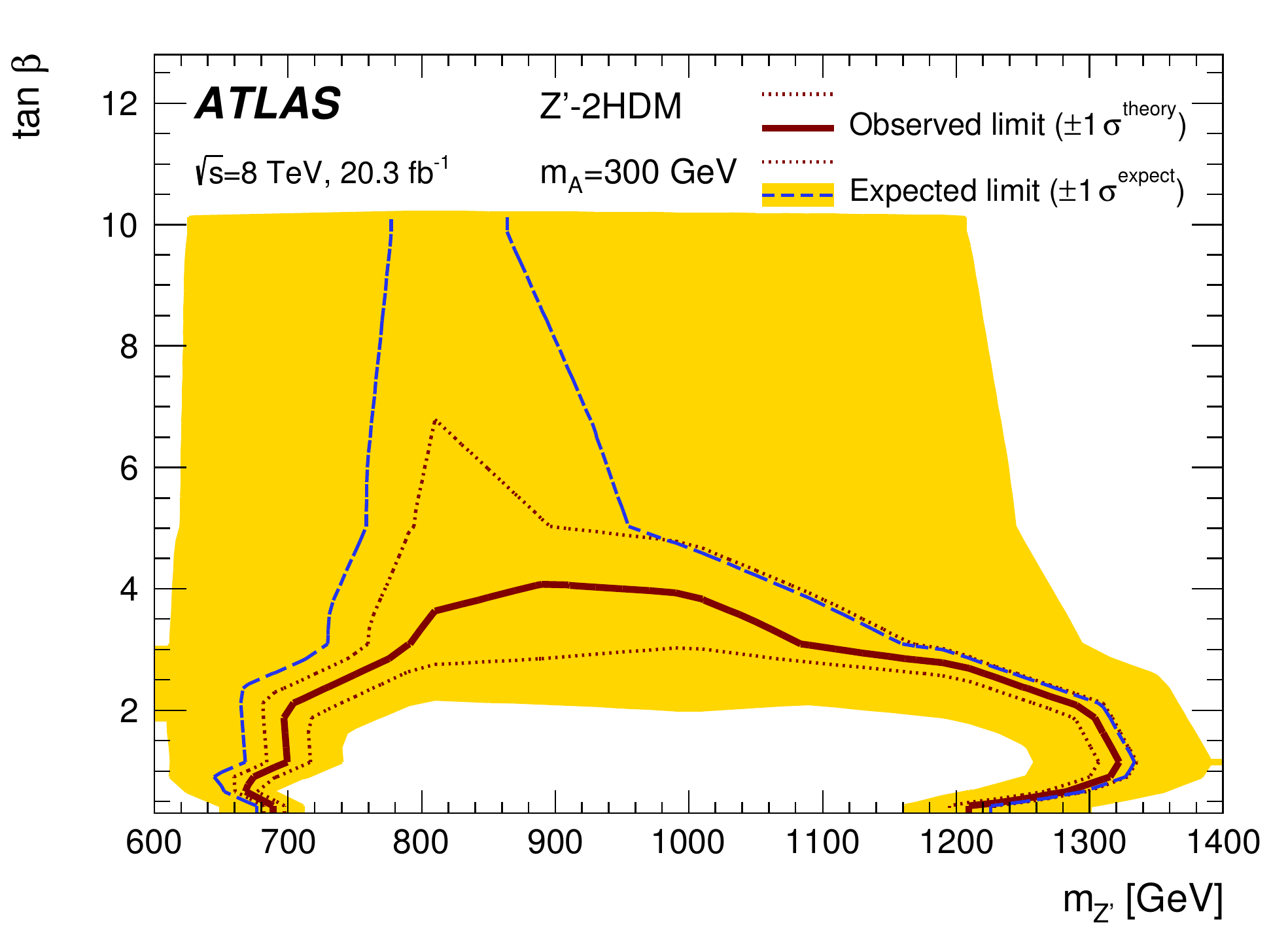}\label{fig:tanbeta}}
\end{center}
\caption{The \Zprime-2HDM exclusion contour in the (a) $m_{\Zprime}$--$m_{A}$ plane for $\tan{\beta} = 1$  and (b) $m_{\Zprime}$--$\tan{\beta}$ plane for $m_{A} = 300$~\GeV. The expected limit is given by the dashed blue line, and the yellow bands indicate its $\pm1 \sigma$ uncertainty.  The observed limit is given by the solid red line, and the red dotted lines show the variations of the observed limit due to a $\pm1 \sigma$ change in the signal theoretical cross-section. The parameter spaces below the limit contours are excluded at $95\%$ CL.}
\label{fig:resexclusion}
\end{figure}

For the resolved channel, the 95\% CL upper limit on the cross-section is derived and used to exclude portions of parameter space of the \Zprime-2HDM model in both the $m_{\Zprime}$--$m_{A}$ and $m_{\Zprime}$--$\tan{\beta}$ planes. In both cases, the \Zprime\ gauge coupling is set to its $95\%$ CL upper limit from precision electroweak constraints and searches for dijet resonances for the corresponding \Zprime~mass and $\tan{\beta}$ value.
Taking the alignment limit of $\alpha = \beta - \pi/2$ evades the constraints in $\tan{\beta}$ for a Type $2$ two-Higgs-doublet model using fits to the observed Higgs boson couplings from the LHC~\cite{Aad:2015pla}. 
The exclusion region in the $m_{\Zprime}$--$m_{A}$ plane is shown in Figure~\ref{fig:resexclusion}\subref{fig:mA}, where $m_{A} \geq 300$~\GeV~in accordance with $b \to s \gamma$ constraints~\cite{Branco:2011iw}.
 For $\tan{\beta}=1$, 
$m_{\Zprime}=700$--$1300$~\GeV~is excluded for $m_{A}$ up to $350$~\GeV, with further exclusion of larger $m_{A}$ for $m_{\Zprime}$ around $1200$~\GeV. Limits in the $m_{\Zprime}$--$\tan{\beta}$ plane are shown in Figure~\ref{fig:resexclusion}\subref{fig:tanbeta}, where $\tan{\beta}$ is $\geq 0.3$ based on the perturbativity requirement of the Higgs--top Yukawa coupling~\cite{Azatov:2012wq}, and is below $10$ based on direct searches for the $A$~\cite{Aad:2014vgg}. For $m_{A}=300$~\GeV, where $A$ decays almost exclusively to a DM pair, $m_{\Zprime}=$ 700--1300~GeV is excluded for $\tan{\beta} < 2$, with further exclusion of larger $\tan{\beta}$ for $m_{\Zprime}$ between $800$~\GeV~and $1000$~\GeV~due to the inclusion of the $\Zprime\to Zh$ contribution in the final state. 
The limits are stronger in regions with larger $m_{\Zprime}$ and smaller $m_{A}$ (or a larger contribution from $\Zprime\to Zh$ where the $Z$ boson is much lighter than $A$), as the harder \MET spectrum in these cases allows a higher \MET requirement with better sensitivity, as demonstrated in Table~\ref{tab:limit_BSM}. The sensitivity eventually drops at very large $m_{\Zprime}$ due to the decrease in signal production cross-section. 

\begin{figure*}[!htbp]
\begin{center}
\begin{tabular}{c@{\hskip 5mm}c}

\subfloat[$\bar \chi \gamma^\mu \chi  B_{\mu\nu} H^\dagger D^\nu H$]{\includegraphics[width=0.495\textwidth]{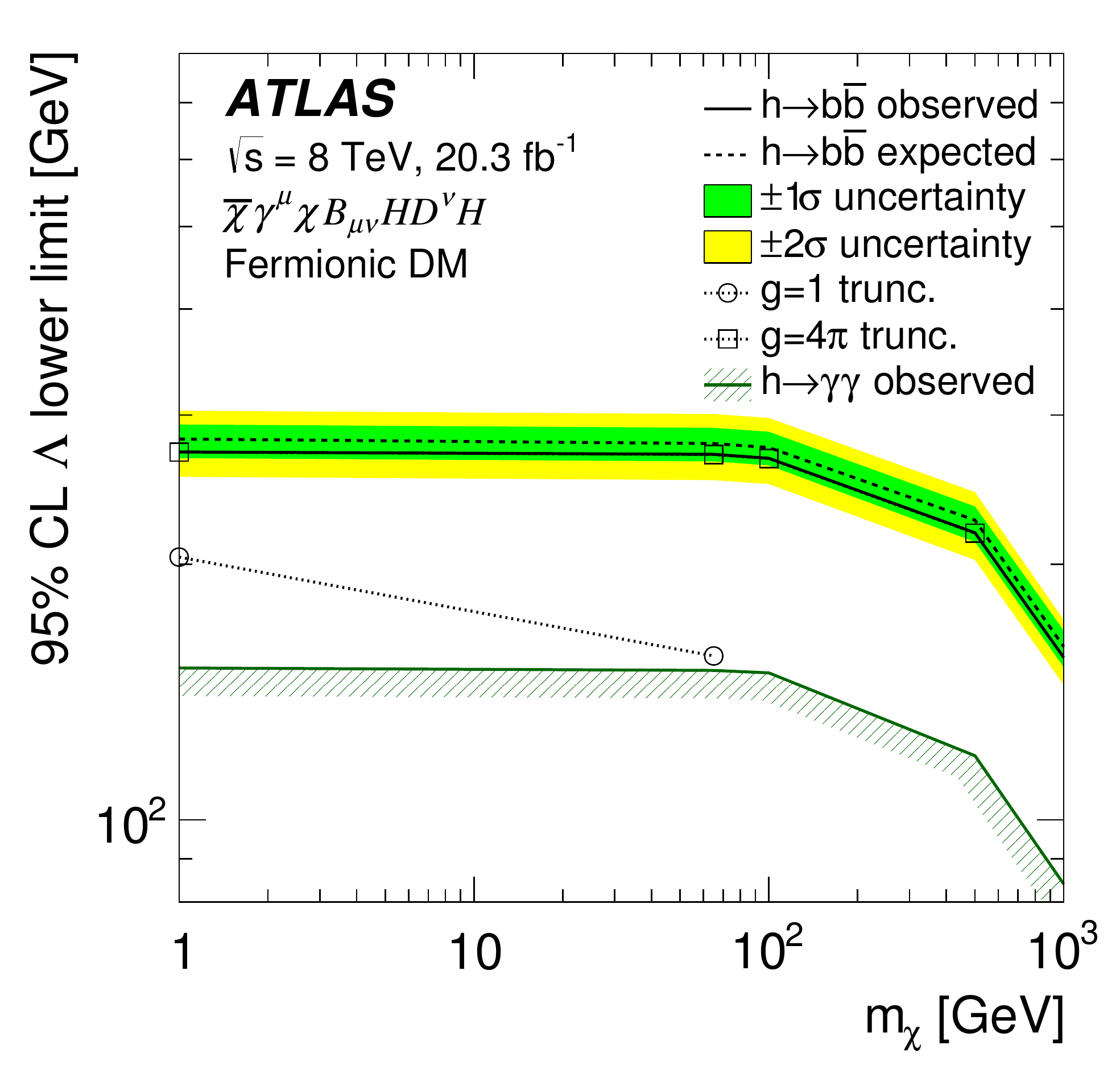}}
\subfloat[$\chi^\dagger \partial^\mu \chi  H^\dagger D_\mu H$]{\includegraphics[width=0.495\textwidth]{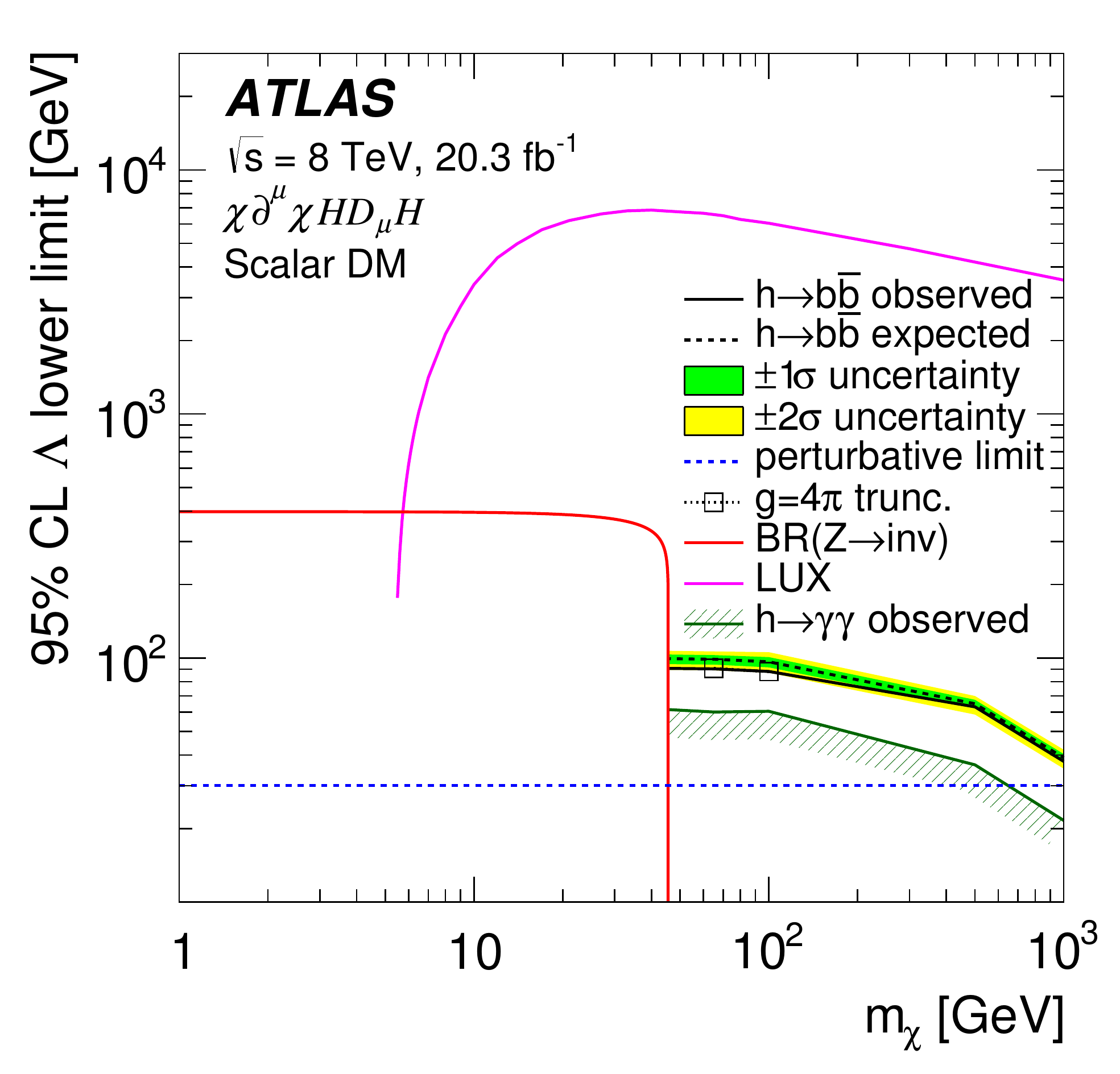}}\\
              
   \end{tabular}
\end{center}
\caption{Limits at 95\% CL on the suppression scale $\Lambda$  as a function of the DM mass ($m_\chi$) for EFT operators (a) $\bar \chi \gamma^\mu \chi  B_{\mu\nu} H^\dagger D^\nu H$ and (b) $\chi^\dagger \partial^\mu \chi  H^\dagger D_\mu H$. Solid black lines are due to $h(\to b\bar{b}) + \MET$ (this article); regions below the lines are excluded. Results where EFT truncation is applied are also shown, assuming coupling values $g=\sqrt{g_qg_\chi} = 1$ (line with circles), $4\pi$ (line with squares). The $g=4\pi$ case overlaps with the no-truncation result. The solid green line with hash marks indicates regions excluded by collider searches for $h(\to\gamma\gamma) + \MET$~\cite{Aad:2015yga}. In the right figure, the region below the dashed blue line fails the perturbativity requirement, the red line indicates regions excluded by upper limits on the invisible branching ratio (BR) of the $Z$ boson~\cite{ALEPH:2005ab}, and the magenta line indicates regions excluded by the LUX Collaboration~\cite{Akerib:2013tjd}.}
\label{fig:eft-limits}
\end{figure*}

For the boosted channel, limits on DM production are derived from the cross-section limits at a given DM mass $m_\chi$, and expressed as 95\% CL limits on the suppression scale $\Lambda$ or coupling parameter $\lambda$ for the effective field theory operators described by Equations~\ref{eq:dim4} to \ref{eq:dim8}. 
As mentioned earlier, the effective field theory model becomes a poor approximation of an ultraviolet-complete model containing a heavy mediator $V$ when the momentum transferred in the interaction, $Q_{\textrm{tr}}$, is comparable to the mass of the intermediate state $m_V = \Lambda\sqrt{g_q g_\chi}$~\cite{Goodman:2010ku,Busoni:2013lha}, where $g_q$ and $g_\chi$ represent the coupling of $V$ to SM and DM particles, respectively. To give an indication of the impact of the unknown ultraviolet details of the theory, a truncation method is adopted~\cite{Busoni:2014sya}, and limits are computed in which only simulated events with $Q_{\textrm{tr}}=m_{\chi\chi} < m_V$ are retained. These limits are calculated for both values of $g=\sqrt{g_qg_\chi}=1$ and $4\pi$, the latter being the maximum possible value for the interaction to remain perturbative. 
The limits are derived assuming that the kinematic properties of the events in the signal processes are independent of $\Lambda$($\lambda$). 
The assumption is not valid in certain regions of parameter space already excluded by invisible Higgs boson~\cite{Belanger:2013xza, Aad:2015pla} or $Z$ boson~\cite{ALEPH:2005ab} decays or near the perturbativity boundary. 
The limits for operators $|\chi|^2|H|^2$ and $\bar{\chi}i\gamma_5\chi|H|^2$ are calculated to be in such regions where the aforementioned kinematic assumption is not valid, hence
only limits for the $\chi^\dagger \partial^\mu \chi  H^\dagger D_\mu H$ and $\bar \chi \gamma^\mu \chi  B_{\mu\nu} H^\dagger D^\nu H$ operators are shown in Figure~\ref{fig:eft-limits} for regions of parameter space where the kinematic assumption holds. 

For both operators shown in Figure~\ref{fig:eft-limits} corresponding to either fermionic or scalar DM candidates, the limits achieved by this analysis are a few times stronger than the prior ATLAS search for DM production in association with a Higgs boson where the Higgs boson decays to a pair of photons~\cite{Aad:2015yga}. 
For the $\chi^\dagger \partial^\mu \chi  H^\dagger D_\mu H$ operator, the $Z$ coupling between DM and nucleon leads to a sizable cross-section for direct detection, and results from the LUX Collaboration~\cite{Akerib:2013tjd} exclude larger regions of parameter space than this search. However, the LUX limits are not applicable if the DM is inelastic leading to insufficient energy transition for direct detection. 
The upper limit on the branching ratio of the $Z$ boson decaying invisibly places stronger constraints for this model for DM with mass values below half of the $Z$ boson mass.  
For the lowest $m_\chi$ region not excluded by results from searches for invisible Higgs boson decays or invisible $Z$ boson decays near $m_\chi=m_{H}/2$, with the kinematic assumption, values of $\Lambda$  up to 24, 91, and 270 GeV are excluded for the  $\bar{\chi}i\gamma_5\chi|H|^2$, $\chi^\dagger \partial^\mu \chi  H^\dagger D_\mu H$, and $\bar \chi \gamma^\mu \chi  B_{\mu\nu} H^\dagger D^\nu H$ operators respectively; values of $\lambda$ above 6.7 are excluded for the $|\chi|^2|H|^2$ operator. 

\FloatBarrier
\section{Conclusion}
\label{sec:conclusion}
A search has been carried out for dark matter pair production in association with a Higgs boson that decays into two $b$-quarks, using $20.3$\,fb$^{-1}$ of $pp$ collisions collected at $\sqrt{s} = 8$~\TeV~by the ATLAS detector at the LHC. Two techniques have been employed, one in which the two $b$-quark jets from the Higgs boson decay are reconstructed separately (resolved), and the other in which they are found inside a single large-radius jet using boosted jet techniques (boosted). A set of increasing \MET thresholds defines the final signal regions for each channel, optimized for individual signals in the parameter space probed. 

The numbers of observed events have been found to be consistent with Standard Model predictions. Results from the resolved channel are used to set constraints in regions of parameter space for a \Zprime-two-Higgs-doublet simplified model. For $m_{A}=300$~\GeV, $m_{\Zprime}=700$--$1300$~\GeV~is excluded for $\tan{\beta} < 2$, with further exclusion of larger $m_{A}$ when $\tan{\beta}=1$. The boosted channel results have been interpreted in the framework of different effective field theory operators that describe the interaction between dark matter particles and the Higgs boson. In addition, model-independent upper limits have been placed in both channels on the visible cross-section of events with large missing transverse momentum and a Higgs boson decaying to two $b$-quarks for each of the ascending \MET thresholds up to \MET $> 400$~\GeV. 

\section*{Acknowledgements}

We thank CERN for the very successful operation of the LHC, as well as the
support staff from our institutions without whom ATLAS could not be
operated efficiently.

We acknowledge the support of ANPCyT, Argentina; YerPhI, Armenia; ARC, Australia; BMWFW and FWF, Austria; ANAS, Azerbaijan; SSTC, Belarus; CNPq and FAPESP, Brazil; NSERC, NRC and CFI, Canada; CERN; CONICYT, Chile; CAS, MOST and NSFC, China; COLCIENCIAS, Colombia; MSMT CR, MPO CR and VSC CR, Czech Republic; DNRF, DNSRC and Lundbeck Foundation, Denmark; IN2P3-CNRS, CEA-DSM/IRFU, France; GNSF, Georgia; BMBF, HGF, and MPG, Germany; GSRT, Greece; RGC, Hong Kong SAR, China; ISF, I-CORE and Benoziyo Center, Israel; INFN, Italy; MEXT and JSPS, Japan; CNRST, Morocco; FOM and NWO, Netherlands; RCN, Norway; MNiSW and NCN, Poland; FCT, Portugal; MNE/IFA, Romania; MES of Russia and NRC KI, Russian Federation; JINR; MESTD, Serbia; MSSR, Slovakia; ARRS and MIZ\v{S}, Slovenia; DST/NRF, South Africa; MINECO, Spain; SRC and Wallenberg Foundation, Sweden; SERI, SNSF and Cantons of Bern and Geneva, Switzerland; MOST, Taiwan; TAEK, Turkey; STFC, United Kingdom; DOE and NSF, United States of America. In addition, individual groups and members have received support from BCKDF, the Canada Council, CANARIE, CRC, Compute Canada, FQRNT, and the Ontario Innovation Trust, Canada; EPLANET, ERC, FP7, Horizon 2020 and Marie Skłodowska-Curie Actions, European Union; Investissements d'Avenir Labex and Idex, ANR, Region Auvergne and Fondation Partager le Savoir, France; DFG and AvH Foundation, Germany; Herakleitos, Thales and Aristeia programmes co-financed by EU-ESF and the Greek NSRF; BSF, GIF and Minerva, Israel; BRF, Norway; the Royal Society and Leverhulme Trust, United Kingdom.

The crucial computing support from all WLCG partners is acknowledged
gratefully, in particular from CERN and the ATLAS Tier-1 facilities at
TRIUMF (Canada), NDGF (Denmark, Norway, Sweden), CC-IN2P3 (France),
KIT/GridKA (Germany), INFN-CNAF (Italy), NL-T1 (Netherlands), PIC (Spain),
ASGC (Taiwan), RAL (UK) and BNL (USA) and in the Tier-2 facilities
worldwide.

\printbibliography

\newpage 
\begin{flushleft}
{\Large The ATLAS Collaboration}

\bigskip

G.~Aad$^\textrm{\scriptsize 85}$,
B.~Abbott$^\textrm{\scriptsize 113}$,
J.~Abdallah$^\textrm{\scriptsize 151}$,
O.~Abdinov$^\textrm{\scriptsize 11}$,
R.~Aben$^\textrm{\scriptsize 107}$,
M.~Abolins$^\textrm{\scriptsize 90}$,
O.S.~AbouZeid$^\textrm{\scriptsize 158}$,
H.~Abramowicz$^\textrm{\scriptsize 153}$,
H.~Abreu$^\textrm{\scriptsize 152}$,
R.~Abreu$^\textrm{\scriptsize 116}$
% ...
\bigskip
\\
% ...
$^{11}$ Institute of Physics, Azerbaijan Academy of Sciences, Baku, Azerbaijan\\
% ...
$^{85}$ CPPM, Aix-Marseille Universit{\'e} and CNRS/IN2P3, Marseille, France\\
% ...
$^{90}$ Department of Physics and Astronomy, Michigan State University, East Lansing MI, United States of America\\
% ...
$^{107}$ Nikhef National Institute for Subatomic Physics and University of Amsterdam, Amsterdam, Netherlands\\
% ...
$^{113}$ Homer L. Dodge Department of Physics and Astronomy, University of Oklahoma, Norman OK, United States of America\\
% ...
$^{116}$ Center for High Energy Physics, University of Oregon, Eugene OR, United States of America\\
% ...
$^{151}$ Institute of Physics, Academia Sinica, Taipei, Taiwan\\
$^{152}$ Department of Physics, Technion: Israel Institute of Technology, Haifa, Israel\\
$^{153}$ Raymond and Beverly Sackler School of Physics and Astronomy, Tel Aviv University, Tel Aviv, Israel\\
% ...
$^{158}$ Department of Physics, University of Toronto, Toronto ON, Canada\\
% ...
\end{flushleft}




\end{document}